\documentclass[12pt,aps,epsfig,amsmath,amssymb]{revtex4}
\usepackage{graphicx}
\usepackage{epsfig}

\newcommand{\Frac}[2]{\frac{\displaystyle #1}{\displaystyle #2}}

\newcommand{\tl}{\tilde}

\begin{document}

\title{\Large\bf
                First normal stress difference and crystallization
                in a dense sheared granular fluid\footnote{
Physics of Fluids, vol. 15, no. 8 (2003)}}
\author{Meheboob Alam$^{1}$\footnote{
Present address: Engineering Mechanics Unit, JNCASR, Bangalore 560064, India;\\
Email: meheboob@jncasr.ac.in}
 and Stefan Luding$^{(1,2)}$}
\address{ $^1$Institut f\"ur Computeranwendungen 1, Universit\"at Stuttgart, 
         Pfaffenwaldring 27, 70569 Stuttgart, Germany \\
         $^2$Particle Technology, DelftChemTech, TU-Delft, \\
         Julianalaan 136, 2628 BL Delft, The Netherlands  %
}
\date{\today}

\vspace*{0.0cm}

\begin{abstract}
The first normal stress difference (${\mathcal N}_1$)
and the microstructure in a dense sheared granular fluid 
of smooth inelastic hard-disks are  probed
using event-driven simulations.
While the anisotropy in the second moment of fluctuation velocity, which is a
Burnett-order effect, is known to be the progenitor
of normal stress differences in {\it dilute} granular fluids,
we show here  that the collisional anisotropies are responsible
for the normal stress behaviour in the {\it dense} limit.
As in the elastic hard-sphere fluids, ${\mathcal N}_1$ remains {\it positive} 
(if the stress is defined in the {\it compressive} sense)
for dilute and moderately dense flows, but  
becomes {\it negative} above a critical density,
depending on the restitution coefficient.
This sign-reversal of ${\mathcal N}_1$ occurs due to the 
{\it microstructural} reorganization of the particles,
which  can be  correlated with
a preferred value of the {\it average} collision angle 
$\theta_{av}=\pi/4 \pm \pi/2$ in the direction opposing the shear.
We also report on the shear-induced {\it crystal}-formation, signalling
the onset of fluid-solid coexistence in dense granular fluids.
Different approaches to take into account the
normal stress differences are discussed in the
framework of the relaxation-type rheological models.
\end{abstract}

\maketitle

\newpage
\section{Introduction}

In the last decade, a lot of  research activity took place
to unveil the properties of granular materials$^{1,2}$,
primarily because of their industrial importance, but also due
to their fascinating properties.
This has unraveled many interesting and so far unresolved
phenomena (for example, clustering, size-segregation, 
avalanches, the coexistence of gas, liquid and solid, etc.).
Under highly  excited conditions, granular materials behave 
as a fluid, with prominent {\it non-Newtonian} properties, like
the normal stress differences$^3$.
While the normal stress differences are of {\it infinitesimal} magnitudes 
in a simple fluid (e.g.\ air and water),
they can be of the order of its isotropic pressure in a 
dilute granular gas$^4$.
From the modelling viewpoint, the presence of such large normal-stress 
differences readily calls for higher-order constitutive models$^{5,6}$
even at the {\it minimal} level.

Studying the non-Newtonian behaviour is itself an important issue,
since the normal stresses are known to be the {\it progenitors} of many 
interesting and unique
flow-features (e.g.\ rod-climbing or Weissenberg-effect, die-swelling, 
secondary flows, etc.$^7$) in non-Newtonian fluids.
Also,  normal stresses can support additional instability modes
(for example, in polymeric fluids and suspensions$^{7-10}$, 
which might, in turn, explain some flow-features of granular fluids.
For example, particle-clustering$^{11-13}$
has recently been explained from the instability-viewpoint
using the standard Newtonian model for the stress tensor$^{12,14,15}$ 

The kinetic theory of Jenkins \& Richman$^{16}$
first showed that the {\it anisotropy} in the second moment of the 
fluctuation velocities, due to the inelasticity of particle collisions, 
is responsible for such normal stress behaviour.
They predicted that the first normal stress difference 
(defined as ${\mathcal N}_1=(\Pi_{xx}-\Pi_{yy})/p$, where $\Pi_{xx}$
and $\Pi_{yy}$ are the streamwise and the transverse components of the 
stress deviator, respectively, and $p$ is the isotropic pressure, 
see section IIB) is maximum in the dilute limit,
decreases in magnitude with density, and eventually approaches zero
in the dense limit. 
Goldhirsch \& Sela$^4$ later showed that the normal 
stress differences appear only at the Burnett-order-description  of 
the Chapman-Enskog expansion of the Boltzmann equation.
Their work has clearly established that the {\it origin} of this effect 
(in the dilute limit) is {\it universal} in both atomic and granular fluids,
with inelasticity playing the role of a {\it magnifier}
and thus making it a sizeable  effect in granular fluids.
While the {\it source} of the normal stress differences in the dilute
limit has been elucidated both theoretically and by simulation, its 
{\it dense} counterpart has not received similar attention so far.
This is an important limit since the {\it onset} of {\it dilatancy} 
(volume expansion due to shear$^{17,18}$), crystallization, etc. 
occur in the dense regime, which in turn would influence
the normal stress differences.

Previous hard-sphere simulations$^{19,3,11}$
did look at the normal stress differences, but they did not probe
the {\it dense} limit in a systematic way.  These simulations showed that
the first normal stress difference vanishes in the dense limit,
in line with the theoretical predictions of Jenkins \& Richman$^{16}$. 
On the other hand, the soft-sphere simulations of 
Walton \& Braun$^{20}$,
with frictional particles, showed that this quantity can change sign
in the same limit.
Our work with smooth inelastic hard-disks unequivocally demonstrates that
${\mathcal N}_1$, indeed, changes its sign at some critical density in the
dense regime, due to the sign-change of its collisional component
at a critical density, which depends on the value of the coefficient 
of restitution ($e$).
More importantly, we show that the origin  of ${\mathcal N}_1$
in the dense limit is distinctly different from
that in a dilute granular gas.
At the {\it microstructural}-level, 
certain topological changes in the {\it anisotropic}
structure of the collision-angle distribution with density are 
responsible for the observed sign-reversal of ${\mathcal N}_1$.

We use the familiar smooth hard-disk model for an event-driven simulation$^{21}$ 
of the uniform shear flow configuration, focussing mainly on
the normal stress behaviour and the microstructure formation as functions
of the density and inelasticity. 
The details of the simulation technique 
and the relevant macroscopic quantities are described in 
section\ \ref{sec:simulation}. 
The simulation results on
the first normal stress difference, the radial distribution function,
the collision angle distribution and the crystalline-structure are presented
in section\ \ref{sec:results}. 
Possible modelling approaches to incorporate  the normal stress differences
are discussed in section\ \ref{sec:nonNewtonian}.
In section\ \ref{sec:conclusion} we summarize our findings, 
with suggestions for possible future work.

\section{Simulation method}
\label{sec:simulation}

We consider a collection of smooth inelastic hard-disks 
in a square box of  size $\tl{L}$ under uniform shear flow --- 
let $\tl{x}$ and $\tl{y}$ be the streamwise and transverse directions, 
respectively, with the origin of the coordinate-frame being positioned 
at the centre of the box. 
The snapshot of a typical simulation, 
with non-dimensional coordinates, is shown in Fig. 1($a$).
Note that the dimensional quantities are denoted by tildes, and 
the reference length, time and velocity scales for non-dimensionalization 
will be specified later in this section.

Let the diameter and the mass of the particle be 
$\tl{\sigma}$ and $\tl{m}$, respectively.
The pre- and post-collisional particle velocities
of particle $1$ are denoted by $\tl{\bf c}_1$ and $\tl{\bf c}'_{1}$, respectively.
Hence, the velocity of particle $2$ relative to $1$ is
$\tl{\bf c}_{21} = \tl{\bf c}_2 - \tl{\bf c}_1$.
Let ${\bf k}_{21}={\bf k}$ be the unit vector directed from the center 
of particle $2$ to that of particle $1$ at contact.
The pre- and post-collisional velocities are related by the expression:
\begin{equation}
 {\bf k}{\bf\cdot}\tl{\bf c}'_{21} = - e({\bf k}{\bf\cdot}\tl{\bf c}_{21}),
\end{equation}
where $e$ is the coefficient of normal restitution,
with $0\le e\le 1$;
note that we restrict ourselves to perfectly smooth particles.
The expression for the collisional impulse is 
\begin{equation}
 \tl{\bf I} \;=\; \tl{m}(\tl{\bf c}'_1 - \tl{\bf c}_1) 
   \;=\; \frac{\tl{m}}{2}(1+e)({\bf k}{\bf\cdot}\tl{\bf c}_{21}){\bf k},
\label{eq:impulse1}
\end{equation}
directed along ${\bf k}$.

\subsection{Model system and algorithm}

The system is periodic in $\tl{x}$-direction, i.e.\ a particle crossing 
the left/right boundary re-enters the system through the opposite 
boundary at the same vertical position with unchanged velocities.
To impose a uniform shear rate ($\tl{\gamma}=\tl{U}/\tl{L}$) in the $\tl{y}$-direction,
the top and bottom image boxes, bounding the central box, are set in motion
with velocities $\tl{U}/2$ and $-\tl{U}/2$, respectively, in the streamwise direction.
This is the standard approach to attain the state of uniform shear flow (USF)
by imparting momemtum transfer by shearing, 
originally introduced by Lees \& Edwards$^{22}$.
Overall, this system represents an {\it extended doubly-periodic} system where the
periodicity in the transverse direction is in the local Lagrangian frame. 
In a typical simulation, the disks are initially placed randomly in the computational box,
and the initial velocity field is composed of the uniform
shear and a small Gaussian random part.
An event-driven algorithm is then used to 
update the system in time, the details of which
may be found in Alam \& Luding$^{23,24}$.

To ascertain whether the system has reached the statistical steady-state, 
the time evolution of the average fluctuation kinetic energy 
(`granular' energy, defined in the next section) is monitored, 
see Fig.\ 1($b$). 
Due to the balance between the
shear work and the collisional dissipation
under homogeneous shear deformation,
the granular energy attains a constant value at the steady state. 
Depending on the value of the 
coefficient of restitution and the number of particles,
it takes about thousand collisions per particle to reach such a statistical
steady-state -- the lower the value of $e$,
the more quickly the system reaches the steady state and {\it vice versa}. 
The simulation is then allowed to run for another $15000$ collisions
per particle to gather data to calculate the macroscopic quantities. 
A few longer runs (30000 collisions per particle) were also checked, 
with no significant change on the measured quantities. 
Another quantity which was simultaneously monitored, along with granular 
energy, is the linearity of the streamwise velocity profile across the 
Couette gap, and we found that the calculated shear rate (i.e.\ 
the slope of the velocity profile) fluctuated around the imposed shear 
rate by at most $1\%$ at densities where crystallization is not evidenced.

\subsection{Macroscopic quantities}
\label{subsec:macro}

With $\tl{L}$, $\tl{\gamma}^{-1}$, $\tl{\gamma}\tl{L}$, 
and $\tl{m}$, used as the
reference length, time, velocity, and mass, 
respectively, the relevant dimensionless
quantities are:
\begin{equation}
   \sigma\;=\; \frac{\tl{\sigma}}{\tl{L}}, \;\;\; 
 ({\bf c}, {\bf u}, {\bf C}) = \frac{1}{\tl{\gamma}\tl{L}}(\tl{\bf c}, 
   \tl{\bf u}, \tl{\bf C}), \;\;\;
   {\bf P} = \frac{{\tl{\bf P}}}{\tl\rho\tl{\sigma}^2\tl{\gamma}^2}, \;\;\;
  T = \frac{\tl{T}}{\tl{\sigma}^2\tl{\gamma}^2}, 
\end{equation}
where ${\bf u}$ is the `hydrodynamic' velocity,
${\bf C} = {\bf c} - {\bf u}$ the fluctuation (peculiar) velocity of
particles, $\tl{\rho}$ the material density of particles, ${\bf P}$ the stress tensor,
and $T$ the granular energy.

The macroscopic stress, as measured in discrete particle simulations, is a 
byproduct of the particle-level mechanisms of momemtum transfer. 
As in the hard-core model of dense gases,
the stress is the sum of its kinetic and collisional components.
The former arises from the transport of momentum 
as the particles move through the system carrying their momentum,
while the latter is due to the direct interparticle collisions.
The homogeneity of the uniform shear
flow allows us to calculate the stress by averaging it 
over the whole computational box$^{3,23,24}$.

The stress, defined in the {\it compressive} sense, 
may be decomposed in the standard way:
\begin{equation}
 {\bf P} = {\bf P}^k + {\bf P}^c 
      =  p{\bf 1} + {\bf \Pi},
\end{equation}
where $p$ is the pressure, ${\bf\Pi}$  the pressure deviator 
and ${\bf 1}$ the unit tensor. 
From the off-diagonal components of the pressure deviator, 
we can calculate 
the {\it shear viscosity} which relates 
the rate of strain to the shear stress:
\begin{equation}
   \mu=  -\Pi_{xy}/\frac{{\rm d}u}{{\rm d}y}.
\label{eq:viscosity1}
\end{equation}
For the steady uniform shear flow, thus, the dimensionless shear viscosity 
can also be interpreted as the shear stress due to our adopted scaling, 
${{\rm d}u}/{{\rm d}y}=\gamma=1$.
The diagonal components of the pressure deviator can be
non-zero, giving rise to normal stress differences.
The first normal stress difference is defined as
\begin{equation}
 {\mathcal N}_1 = \frac{({\Pi}_{xx}-{\Pi}_{yy})}{p}.
\label{eq:nsd1}
\end{equation}
Note that we have scaled this quantity by pressure to ascertain
its relative magnitude with respect to pressure.
For a standard Newtonian fluid, ${\mathcal N}_1=0$ 
and thus ${\mathcal N}_1$ is  an indicator of 
the {\it non-Newtonian} character of the fluid.
${\mathcal N}_1$ can be decomposed into 
kinetic and collisional parts:
\begin{equation}
 {\mathcal N}_1 = {\mathcal N}_1^k + {\mathcal N}_1^c
 =\frac{(\Pi_{xx}^k-\Pi_{yy}^k)}{p}
 +\frac{(\Pi_{xx}^c-\Pi_{yy}^c)}{p}.
\label{eq:nsd2}
\end{equation}

Note that the
{\it sign} of ${\mathcal N}_1$ crucially depends
on the convention used to define the stress tensor.
For example, in the rheology literature, stress is typically
defined in the {\it tensile} sense$^7$.
A positive  ${\mathcal N}_1$ for the compressive case 
is equivalent to its negative value for the tensile case
and vice versa. This point should be kept in mind
while making any comparison with data in the rheology literature.

From the trace of the kinetic stress tensor, ${\bf P}^k$,
one can calculate the granular energy,
\begin{equation}
  T = \frac{1}{2\sigma^2}\left[\frac{1}{N}\sum_{i=1}^{N}{C}_i{C}_i \right],
\label{eq:genergy1} 
\end{equation}
which is a measure of the random motion of the particles with respect 
to the mean motion. 

There are two dimensionless control parameters:
the volume fraction of particles ($\nu$) and  
the coefficient of normal restitution ($e$). 
The shear rate is also a control parameter, however, due to 
normalization we have $\gamma=\tl{\gamma}/\tl{\gamma}=1$,
and changing the value of $\tl{\gamma}$ does not influence the reported 
results; the imposed shear rate $\gamma$ is thus kept fixed at unity.
The simulations are carried out for the whole range of
solid volume fractions, varying from the dilute to the dense limit,
over a large range of values for the coefficient of 
restitution ($e=0.3$--$0.99$).
For most of the simulations, the number of particles
are fixed to $N=1024$, and increasing the value of $N$ by fourfold ($N=4096$)
did not affect the reported quantities noticeably;
for example, the change in ${\mathcal N}_1$ was about
$4.3\%$ and $5.1\%$ at $\nu=0.3$ and $0.75$, respectively,
for a restitution coefficient of $e=0.9$.
We note here that the system-size dependence of the
rheological quantities (pressure and viscosity) is known to be strong only for
a small number of particles ($N<100$)$^{11,24,25}$.

For the typical simulation in Fig. 1($a$), at steady-state,
after $2 \times 10^7$ collisions, the parameter values were
$\nu=0.5$, $N=1024$ and $e=0.7$.
The variations of the granular energy $T$ and the calculated shear
rate $\gamma_{cal}$ with time are shown in Fig. 1($b$),
along with corresponding initial variations in two insets.
(The data represent the instantaneous values of $T$ and $\gamma_{cal}$ 
sampled at a regular interval of $400$ collisions -- no time averaging is involved here.)
Note that $\gamma_{cal}$ was computed by binning the system
into $20$ equal-size bins in the transverse direction
and taking averages over all particles in each bin.
It is observed that the granular energy reaches
its steady value ($T=0.6121 \pm 0.022$) quickly after the initial transients
and the calculated shear rate fluctuates around its imposed value ($\gamma=1$)
by about $\pm 1\%$. The fluctuations in both $T$ and $\gamma_{cal}$
at steady-state are due to the finite-size of the system and
diminish with increasing  number of particles as $N^{-1/2}$.

\section{Results}
\label{sec:results}

For detailed results on the quantities pressure, shear viscosity and 
granular energy, and for their comparison with kinetic theory predictions,
we refer to our recent study$^{24}$.
Here, we will mainly focus on  the behaviour of the first
normal stress difference and its kinetic and collisional components.
We also present results on the pair distribution function and the
collision angle distribution to characterize microstructures.
Lastly we will present results on {\it crystal}-formation at 
high densities, signalling the coexistence of fluid and solid,
complementing recent results in non-sheared systems$^{26,27}$.

\subsection{Normal stress difference}
\label{subsec:nsd}

Figure 2($a$) shows the variation of the first normal stress
difference (${\mathcal N}_1$) with density for two values
of the coefficient of restitution.
It is observed that ${\mathcal N}_1$ is maximum at the
dilute limit and decreases thereafter with $\nu$.
The overall variation of ${\mathcal N}_1$ with density 
looks similar at other values of the coefficient of restitution, with a
difference in the magnitude of ${\mathcal N}_1$.
The inset in Fig. 2($a$) shows that
${\mathcal N}_1$ decreases quite sharply in the dense limit
and becomes negative at some density ($\nu=\overline\nu$).
Increasing the value of the restitution coefficient decreases
this critical density $\overline\nu$.
The arrows on the left-ordinate indicate the asymptotic values of 
${\mathcal N}_1$ for a two-dimensional granular gas in the dilute limit$^4$:  
\begin{equation}
 {\mathcal N}_1  = 1.0448(1-e^2) . 
\label{eq:nsd_jen1}
\end{equation}
The anisotropy in the second moment of the fluctuation velocity
is {\it primarily} responsible for the finite  normal stress difference 
in the dilute limit$^{16,28}$
and this  shows up only at the Burnett-order of the Chapman-Enskog 
expansion$^4$.
We should mention here that 
the limit $e\to 1$ is singular and the normal stress
difference survives even in the elastic limit as pointed
out by Goldhirsch \& Sela$^4$.
The corresponding expression
for ${\mathcal N}_1$ in a molecular gas is:
\[
 {\mathcal N}_1 \approx 1.358 \frac{\gamma^2 {\ell}^2}{\langle u^2\rangle},
\]
where $\ell$ is the mean free path and $\langle u^2\rangle$ is the
rms of the velocity fluctuations.
However, because of its extremely small magnitude under normal conditions,
the normal stress difference is not {\it measurable} in 
a molecular fluid.

Previous hard-sphere simulations of Campbell
and coworkers$^{3,19}$ are in {\it variance} with our result
in that they found ${\mathcal N}_1\to 0$ as $\nu\to\nu_{max}$.
However, the soft-sphere simulations of Walton \& Braun$^{20}$,
with frictional particles, support our observation
that ${\mathcal N}_1$ indeed undergoes a {\it sign-reversal}.
To better understand what is responsible for
the sign-reversal of ${\mathcal N}_1$, we look at the kinetic and
collisional components of the first normal stress difference.
Figure 2($b$) shows the variations of ${\mathcal N}_1^k$ and
${\mathcal N}_1^c$ with density at $e=0.7$.
We observe that ${\mathcal N}_1^k$ is maximum at the
dilute limit and decreases monotonically 
to zero as $\nu$ approaches the packing limit.
Except for the dense limit, the overall behaviour
of ${\mathcal N}_1^k$ represents that of the
total normal stress difference.
The collisional component, ${\mathcal N}_1^c$, 
shows a {\it non-monotonic} variation with density:
${\mathcal N}_1^c$ is zero in the dilute limit,
increases with increasing $\nu$, 
remains almost constant for intermediate densities,
and then decays sharply in the dense limit. Interestingly,
${\mathcal N}_1^c$ becomes negative at some 
critical density ($\nu=\overline\nu$) 
beyond which the behaviour of ${\mathcal N}_1^c$
mirrors that of ${\mathcal N}_1$ (see inset).
Thus the normal stress behaviour in the dense regime  is clearly
due to the anisotropy in the collisional stress.

Recall that the kinetic theory of Jenkins \& Richman$^{16,28}$
predicts that ${\mathcal N}_1\to 0$ in the dense limit.
The predictions of the revised Enskog theory of Santos {\it et al.}$^{29}$
are in line with that of Jenkins \& Richman, even though
their kinetic model is claimed to be valid
even in the crystalline-phase. 
Since the source of normal stress differences in all
these theories is linked to the anisotropy in the second moment
of velocity fluctuations (which vanishes as $\nu\to\nu_{max}$),
they are unable to predict the correct behaviour of normal
stresses in the dense limit.
We would like to stress here that, 
as mentioned in the Introduction, many fascinating
non-Newtonian effects$^7$ are primarily determined by the
first normal stress difference and its sign. For example, the
rod-climbing effect can occur in a granular fluid if 
${\mathcal N}_1 < 0$, {\it i.e.} only in the dense limit
(assuming that the second normal stress difference is negligible).
Furthermore, having a constitutive model which
reproduces the correct sign of ${\mathcal N}_1$
is also important, since it is well known in the
rheology literature that the {\it odd} sign for ${\mathcal N}_1$
leads to the {\it instability} of the rest state$^{30}$.

The sign-reversal of ${\mathcal N}_1^c$ can be succinctly presented 
as a phase-diagram in the ($\nu, e$)-plane by plotting the
zeros of ${\mathcal N}_1^c$ as a function of 
the coefficient of restitution, see Fig. 3.
Below the solid line,
${\mathcal N}_1^c$ is positive, and negative above it.
Also plotted in this figure is the line for the zeros of ${\mathcal N}_1$
which, as expected, lies slightly above.
Thus, the first normal stress difference is zero along the solid line
which may be called the {\it symmetry-line}.
It is observed that 
decreasing the coefficient of restitution increases
the critical density ($\overline\nu$) at which
${\mathcal N}_1^c$ changes sign. 
As we approach the elastic limit, $\overline\nu$
depends strongly on the value of $e$.
We further note that as $e\to 1$, $\overline\nu\sim 0.62$
which is well below the freezing-point density
of a 2D hard-disk fluid, $\nu_f \approx 0.70$$^{26,27}$.

We need to mention here that the {\it sign-change} of ${\mathcal N}_1$
is not uncommon in other non-Newtonian fluids. For example,
in non-Brownian viscous suspensions,
${\mathcal N}_1$ changes sign at high Peclet number$^{31,32}$.
However, the reason for this effect is quite different
in granular fluids as we show below.

\subsection{Microstructural features}

To understand the microstructural mechanism for the origin of the first
normal stress difference and its sign-reversal,
here we probe several microstructural features of a dense granular fluid.

\subsubsection{Radial distribution function}

Typical snapshots of the system in the dense regime are shown 
in Fig. 4 at four different densities with $e=0.7$.
Note that the density for the subplot 4($c$) is $\nu=0.725$ 
for which ${\mathcal N}_1^c\approx 0$.
Looking at the corresponding distribution of granular energies
(not shown here for brevity),
we could find signatures of clusters (group of particles)
with lower energies surrounded by particles with higher energy.
To understand the 
flow-microstructures and their energetics at such high densities,
we need to probe the pair distribution function and similar measures
for the structure of the packing.

Figure 5($a$-$d$) shows the radial distribution function $g(r)$
at four densities, with parameter values as in Fig. 4.
The thin, dotted lines are data from a non-sheared, homogeneous,
elastic system$^{26,27}$,
whereas the thick, solid lines
represent a sheared situation with rather strong dissipation $e=0.7$.
(Note that the elastic distribution function has been measured
under non-sheared periodic boundary conditions; also for $e$ somewhat
smaller than unity,  the same results were obtained 
as long as the system remains homogeneous.)
We observe that the weak difference at low density $\nu=0.6$ grows with increasing
density, concerning two aspects:\\
(i) The peak value of contact in the
sheared systems is always larger than that in a homogeneous system of the same
density, and the difference increases strongly with density, 
another indicator for clustering$^{33}$.\\
(ii) The peaks and valleys, which allow to distinguish between different
lattice structures, are different in the sheared case when compared to
the non-sheared situation.  In the former case, peaks at 
$r/\sigma=1,2,3,\ldots$ are observed,
signalling shell-formation about any test-particle.  In the latter
case, the peaks at $r/\sigma=1,\sqrt{3},2,\ldots$ indicate a crystallization
transition and the development of a triangular lattice. 

The peaks in the sheared situation, e.g.\ at $r/\sigma=2,3$, become 
sharper as the density is increased, but the one at $r/\sigma=4$
is not well-defined even at $\nu=0.75$; higher order peaks are almost
invisible, indicating long-range disorder due to the shearing in contrast
to the long-range order that evolves in the non-sheared system.
The comparison between the sheared and non-sheared cases
suggest that the structure-formation is much slower in a sheared fluid
which, in turn, implies that, as expected, the freezing-point density of the former 
would be larger than that of the latter.  
The splitting of the second-peak in the non-sheared case
corresponds to the {\it onset} of freezing transition$^{34}$;
with increasing density this splitting becomes much more prominent,
with similar structural-features appearing at the successively 
higher-order peaks. 
For the sheared system, however, we do not observe
similar splitting, rather we see a {\it sharp} second-peak.
This also occurs for a highly inelastic sheared fluid at 
a much lower density$^{33}$; hence, a sharp second peak
is a signature of short-range ordering due to 
the dissipative particle-clustering.
The higher-order peaks in the sheared case become 
prominent only if we go beyond $\nu=\pi/4$, which corresponds to
the limit of perfect square-packing, but it is difficult to maintain 
homogeneous-shearing at such high densities, because the system
splits into two parts, a dense, cold, crystalline area and a dilute,
hot, fluid area -- see below for details.

Figure 6 shows the variations of the pressure
and viscosity functions ($f_p=p/\rho T$ and $f_\mu=\mu/\rho\sigma\sqrt{T}$, respectively)
with density at a restitution coefficient $e=0.7$.
We observe that both increase {\it monotonically} 
with density, much beyond $\nu=0.7001$, and we did not find the 
{\it hysteretic} van-der-Walls loop in our pressure data upto $\nu=\pi/4$,
another indication that the crystallization is hindered/delayed by shear.
These observations, together with our result 
that in the elastic limit ${\mathcal N}_1$ changes sign at a
much lower density ($\nu\approx 0.62$, see Fig. 3) 
than the corresponding freezing-density, suggest that the sign-reversal
of ${\mathcal N}_1$ is not related to the freezing-transition.

\subsubsection{Collision angle distribution}

Associated with the {\it sign-reversal}  of the 
first normal stress difference is a change
in the relative magnitudes of the normal
stress components ($P_{xx}$ and $P_{yy}$).
Subtle changes in the direction and magnitudes of the collisional-mode 
of momemtum transfer could  influence
the individual components of the stress tensor.
In order to  test this hypothesis,  we focus
on the collision angle distribution function, $C(\theta)$,
which is defined  such that $C(\theta)d\theta$ is the
probability of collisions occuring at an angle 
lying between $\theta$ and $\theta+d\theta$, 
with the angle $\theta$ being measured in the anticlockwise direction
from the positive $x$-axis (see Fig. 13).
For a fluid in equilibrium, all collisions
are equally likely,  and hence $C(\theta)$ is a uniform function of $\theta$,
{\it i.e.} $C(\theta)=1/\pi \approx 0.318309$.
For a non-equilibrium system (e.g. shear flow), however,
preferred collisions are dictated by the nature of the external field,
leading to an anisotropic distribution for $C(\theta)$$^{35-38}$.
Following Savage \& Jeffrey$^{36}$ and Campbell \& Brennen$^{37}$,
an explicit expression for $C(\theta; \nu, e)$ can be derived 
for the case of uniform shear flow as detailed in the Appendix.
Note that the angular dependence of $C(\theta;\nu,e)$ will be modified
by both the density and the restitution coefficient.

Figure 7 shows the comparison of our
simulation data on $C(\theta)$ with the
theoretical predictions of equation (\ref{eqn_colang1})
for two different values of the restitution coefficient 
at a density $\nu=0.6$.
It is  observed that the probability
of collisions is higher  on the {\it upstream}-faces of
the colliding particles, {i.e.} for  $\theta\in [\pi/2, \pi]$
and $\theta\in [-\pi/2,0]$ (i.e. the hatched-areas on the test-particle in Fig. 13).
This is a consequence of the imposed shear-field which
compresses  the flow-structure along the $3\pi/4$-direction and
stretches it along the $\pi/4$-direction. 
Regarding the comparison with theory, 
there is, clearly, a phase-difference between theory and simulation,
and the overall agreement is only qualitative.

Note in Fig.\ 7 that the probability of collisions on
the upstream-faces  increases further 
as the restitution coefficient increases.
This, in turn, suggests that particle motion becomes more
{\it streamlined} (i.e. along the streamwise direction)
with increasing dissipation levels, which will naturally lead to a
reduction in the transverse component 
of the fluctuation velocities of the particles.
Thus, the {\it macroscopic} manifestation of
such {\it microscopic} streamlined-motion would
be an increase in the magnitude of the kinetic component of the first normal
stress difference (${\mathcal N}_1^k$).

Turning our attention to the range of densities where 
${\mathcal N}_1$ undergoes a sign-reversal,
we show the collision angle distributions $C(\theta)$ in Fig. 8
as polar plots with $e=0.7$;
the corresponding densities are as in the subplots of Fig. 4.
It is observed that the {\it anisotropic} structure of $C(\theta)$ gets
further modified in this regime, with 
distinct peaks appearing near $\theta=0$ and $2\pi/3$
(see subplot $b$). While the peak at $\theta=0$ corresponds
to {\it head-on} collisions between particles in the same-layer,
the one at $\theta=2\pi/3$ clearly signals the {\it onset}
of triangular-structure formation.
Another noteworthy point is that
the collisions on the downstream-faces of the colliding particles
are {\it rare} at these densities and hence $C(\theta)$ can be 
approximated solely by its contributions from the second- and 
fourth-quadrants ($\theta\in [\pi/2, \pi]$ and $\theta\in [-\pi/2, 0]$, 
respectively).

Since the momentum transfer occurs mainly due to 
collisions in the dense regime,
the stress tensor can be approximated by
\[
 {\bf P} \sim \int ({\bf k}\otimes{\bf k})C(\theta){\rm d}\theta,
\]
where ${\bf k}$ is the unit vector joining the line of centres
of the two colliding disks.
Assuming now that all the collisions would occur at some average
collision angle $\theta_{av}$ so that $C(\theta)=C(\theta_{av})$,
and recalling that $C(\theta)$ is 
well represented in this regime by  restricting $\theta$ only in the second-
and fourth-quadrants,
the expression for the first normal stress difference
simplifies significantly to
\begin{equation}
 {\mathcal N}_1 \sim \left[ ({\bf k}\otimes{\bf k})_x - 
     ({\bf k}\otimes{\bf k})_y\right]_{\theta=\theta_{av}} C(\theta_{av}).
\end{equation}
It is trivial to check that ${\mathcal N}_1=0$ at $\theta_{av}=-\pi/4$.
From our simulation data, we have calculated $\theta_{av}$
by averaging $C(\theta)$ over the second- and fourth-quadrants,   
whose variation with density is plotted  in Fig. 9  
for two restitution coefficients. It is observed
that $\theta_{av}$ crosses  through $-\pi/4$ (i.e. $3\pi/4$) at around  
the critical density $\overline{\nu}$ for all restitution coefficients.
For example, $\theta_{av}\approx -45.16^{\circ}$ and $-45.04^{\circ}$ 
at $\overline{\nu}=0.725$ and $0.67$, respectively,
where ${\mathcal N}_1$ changes sign.
Thus, the microstructural signature of the sign-reversal of ${\mathcal N}_1$
is directly correlated with the {\it average} collision angle being greater 
or less than $-\pi/4$ (or $3\pi/4$).

\subsection{Crystallization: Fluid-solid coexistence}

Figure 10($a$) shows a snapshot of the system at the steady state
with parameter values being set to $\nu=0.8$ and $e=0.9$.
It is observed that  a solid-layer coexists
with two fluidized zones on either side of it.
A closer look into the solid-layer
reveals that the particles are arranged in a triangular-packing,
representing a {\it crystal}, 
and thus we have a clear evidence for {\it fluid-solid} coexistence. 
The corresponding {\it instantaneous} streamwise velocity  
profile at $t=390$ (i.e. the image-boxes have moved $390$ strain units
from their original position) is shown in Fig. 10($b$); 
the coarse-graining is done by binning the system into 20 equal-size bins 
in the transverse direction and then 
taking averages over all the particles in each bin.
Clearly, the shearing is {\it inhomogeneous} across the Couette-gap:
the crystal is aligned along the streamwise direction
and hence we call it a {\it layered}-crystal;
the shear-rate in the fluidized regimes on 
either side of the crystal is almost uniform.
Note that the {\it asymmetric} nature of the velocity profile  
also signals the breakdown of the Lees-Edwards boundary condition
as a motor for the homogeneous shear. 
The formation and the time-evolution of this crystal can be
ascertained from Fig. 10($c$) which shows the corresponding
evolution of the streamwise velocity at early times.
We observe that the crystal has fully formed at $t=76$, and
the velocity profile remains antisymmetric about $y=0$ till $t=150$.
With further time-evolution, however, the crystal does not remain stationary
in the transverse direction, rather it moves slowly
with particles diffusing across the fluid-solid interface.
The overall life-time of this crystal is several orders of magnitude 
larger than the external time-scale $\gamma^{-1}$, imposed by the shear.
The corresponding collision-angle distribution $C(\theta)$ 
in Fig. 10($d$) shows three distinct peaks at $\theta=0$, $\pi/3$
and $2\pi/3$. Note that the peak at $\theta=\pi/3$
does not exist in the fluid phase (refer to Figs. 8$a$), and 
this provides evidence that
the particles in the crystalline-phase are arranged in 
the triangular-packing structure.
The large area of the crystalline phase
as well as its relatively large value of solids fraction
(hence, small mean free time) further suggest
that most of the collisions occurred in the crystalline-phase.

We should remark here that, at such high densities, the 
{\it inelastic collapse}$^{39-41}$ would eventually terminate
the evolution of the system.
We used the TC-model$^{40}$ to avoid inelastic collapse,
but could not altogether eliminate it within the crystal after some time.
But the important point to note is that the crystalline-phase
can be maintained for a long period of time ($t>100$ strain units),
and hence the reported results are not transient effects.

Analogous plots for a nearly elastic system ($e=0.99$) are shown in 
Fig. 11($a$--$c$) for the same density $\nu=0.8$.
The overall features are similar to that for $e=0.9$, but the width of the
crystalline zone is a little larger.
By decreasing the dissipation-level to $e=0.7$, we did not observe
crystal formation, with other parameters being fixed; by increasing
the system-size to $N=4096$, however, we observed layered crystal at $e=0.7$.
Thus, the formation of such layered-crystalline structure depends 
crucially on the system size and the dissipation level:
{\it the larger the system-size and/or the weaker the dissipation, 
the more susceptible the system is to crystallize}.

Note that even if we are well below the limit of
perfect square-packing ($\nu_{sp}=\pi/6\approx 0.785$),
the system could crystallize if the dissipation-levels are low;
for example, we observed layered crystalline structures
at $\nu=0.75$ with $e=0.99$ and $N=4096$.
Decreasing the coefficient of restitution to $e=0.9$,
the flow-field remained homogeneous.
Thus, our layered crystalline structures
appear to be tied to a {\it long-wave} instability of 
the elastic hard-sphere fluids$^{56}$.
Since our results are not driven by the inelastic
dissipation, they are distinctly different from
the layered shear-banding patterns, 
as predicted by the kinetic theory models$^{15,42,43}$,
in granular Couette flows.
Having said that we need to mention that
such {\it dissipative} layering patterns (i.e. those which become stronger
with increasing dissipation levels$^{15}$) were also
found in simulations of a dilute sheared granular fluid$^{43}$,
but the simulations were allowed to evolve from
an {\it unsheared} initial configuration with a special kind of boundary condition.
In contrast, all our simulations started from a uniform
shear condition.
In the present contribution, we have mainly focussed  on 
the non-dissipative layering in the dense limit, 
and the related issues of dissipative layering are
relegated to a future study.

Before moving further, we make a qualitative  comparison with the
earlier simulation work of Campbell \& Brennen$^{37}$ 
on the bounded Couette flow of inelastic, frictional hard disks.
They also reported similar layered-microstructure but due to the
small system-size ($N=40$) and boundary-effects (they considered
a shear-flow bounded by frictional walls), 
the distinct shear-band formation that we have reported is
not evident in the snapshots of their simulations.
Nevertheless, we believe
that our results are akin to that reported by Campbell \& Brennen.
One of the referees has drawn our attention to the recent
work of Campbell$^{44}$ who probed the dense limit 
of a three-dimensional Couette shear flow using soft-sphere simulations.
He was able to maintain uniform shearing even at
a density of $\nu=0.62$, but beyond that he reported
shear-band formation in that the flow-field
degenerates into sheared and non-sheared zones in the gradient direction.
This simply suggests that one
can maintain uniform shearing in a three-dimensional geometry even beyond
the analogous square-packing limit ($\phi=\pi/4\approx 0.52$)
since the particles have now an additional degree of freedom,
orthogonal to the shear plane, to rearrange themselves.
Moreover, most of Campbell's  simulations were done with $1000$ particles
(equivalently, $100$ particles in 2D) at a restitution
coefficient of $0.7$. With these parameter values, we did 
not find layered crystals at $\nu=0.8$ in two-dimensions
(see discussion in the next section).

\subsubsection{Shear-induced ordering and Reynold's dilatancy}

One of the referees has drawn our attention to the recent work of Lutsko$^{45}$
who studied the shear-induced ordering in a low-density ($\nu\approx 0.26$) 
elastic hard-sphere fluid.
The earliest simulations of Erpenbeck$^{46}$ on 3D
elastic hard-sphere fluids showed that at high shear rates
the system breaks down into orderded (solid) and disordered (fluid)
phases in the direction of the mean vorticity (i.e. normal to the shear-plane).
This induces a  long-range two-dimensional ordering, called a {\it string-phase},
in the shear-plane. More importantly, such ordering occurs only
for a range of shear-rates-- the lower the density, the larger this shear-rate interval.
However, the latter work of Evans and Morriss$^{47}$ showed that
the disorder-order transition of Erpenbeck 
arises due to the profile-biased-thermostat$^{47}$ since the string-phase vanishes
completely when a profile-unbiased-thermostat$^{47}$ is used.
Thus, the shear-induced-ordering in elastic hard-sphere fluids
appears to depend on the choice of the thermostat. 
Note further that all the above works probed moderately
dense systems only, well below the corresponding square-packing limit.
Now to compare these results
with our observations in granular fluids, we first
need to define an equivalent shear rate since the
granular energy and the shear rate are dependent on each other for the latter system.
From the energy balance equation, it is trivial to show that
the granular energy has the following functional relation
with the shear rate and the restitution coefficient:
\[
 T \;\;\; \propto\;\; \frac{\gamma^2}{1-e^2}\;\; \propto \;\;   {\gamma^*}^2,
\]
where $\gamma^*=\gamma/(1-e^2)^{1/2}$ 
is defined as the {\it reduced shear rate}.
In the quasielastic limit ($e\to 1$) the
reduced shear rate (and hence the granular energy) approaches infinity, 
which is equivalent to the high shear-rate limit of
an elastic hard-sphere fluid.
Since we observed layered crystalline structures
only in the quasielastic limit,
we may thus conclude that such structures would also persist in 
elastic hard-disk fluids at large shear rates.

It is interesting to ask whether the layered crystalline-structures
of Figs. 10 and 11 are, in any way, related to the concept of
the Reynold's dilatancy$^{17,18}$
which is explained schematically in Fig. 12.
The top two sketches depict the classical {\it constant-load}$^{48}$ 
shear-cell experiment (in two-dimensions) in which the material is subjected to
a constant normal load.
Fig. 12($a$) corresponds to an ideal situation of maximum packing, 
with the particles being arranged in a triangular lattice ($\nu=\nu_{tp}=\pi/2\sqrt{3}$).
Clearly, in this situation the top and the bottom plates
of the shear cell will simply slide over the material, without deforming it.
However, if one of the plates is allowed to move in the
vertical direction and thereby allowing the particles to rearrange
themselves, e.g. as in Fig. 12($b$), 
the material can be deformed even homogeneously (if $\nu\leq \nu_{sp}=\pi/6$). 
This is the {\it shear-coupled
volume change}, commonly known as the Reynold's dilatancy.
Note, however, that our simulations mimic {\it constant-volume} experiments,
since the volume of the computational box remains fixed (see Figs. 12$c$ and 12$d$);
but, of course, now the pressure (i.e. normal load) 
can vary in response to particle motions inside the shear cell.
For this case also, the uniform shearing is possible if and only if the overall density 
remains below the square-packing limit ($\nu <\nu_{sp}$).
However, for densities above this value ($\nu_{sp}< \nu < \nu_{tp}$),
the shearing can be started  only if we allow the particles
to rearrange themselves. This is possible if a part of the
system becomes denser, allowing {\it free} volumes to the rest of the system
which is nothing but Reynold's dilatancy too.
Hence, we will  end up with a crystalline-phase coexisting with
a fluid phase, a typical example of which is shown schematically in Fig. 12($d$).
Thus the phenomenon of Reynold's dilatancy,
for densities $\nu_{sp}< \nu < \nu_{tp}$,
would make the {\it ordering} transition, as depicted
in Figs. 10 and 11, more prominent, 
and this effect would be much stronger in two-dimensions.

\section{Consequences for the constitutive modeling: Relaxation models}
\label{sec:nonNewtonian}

Here we attempt to describe the normal-stress behaviour of a granular fluid
using the standard relaxation-type  models. 
Prior literature on the dense-gas kinetic theory, which forms the
foundation of theoretical developments of granular fluids
in the {\it rapid-shear} regime, indicates that such a
stress relaxation mechanism does also exist in granular fluids$^{49-51}$.
The relaxation-type models are routinely used to describe the 
non-Newtonian behaviour of viscoelastic/viscoplastic materials,
and hence might be apt for granular fluids in the dense limit as well.
The recent work of Zhang \& Rauenzahn$^{52-53}$ 
clearly shows that such viscoelastic stress
relaxation mechanism exists in dense granular flows.
Following a rigorous statistical mechanical procedure,
they derived an evolution equation for the collisional stress
tensor which boils down to a frame-indifferent viscoelastic model, 
with the Jaumann derivative
appearing directly without appealing to objectivity arguments.

Let us consider the viscoelastic relaxation approximation
suggested recently by Jin \& Slemrod$^{51}$
to regularize the Burnett order equations of Sela \& Goldhirsch$^5$
for a low-density granular fluid.
Their proposed equation for the pressure 
deviator, in our notation, is
\begin{eqnarray}
 & & {\bf\Pi} + \tau_1\left(\frac{D{\bf\Pi}}{Dt}-{\bf L}^T\cdot{\bf\Pi}
  -{\bf\Pi}\cdot{\bf L}
   + {\textstyle\frac{2}{d}}tr({\bf\Pi}\cdot{\bf L}){\bf 1}\right) \nonumber\\
 & & \hspace*{3.0cm} 
 +\;\; \tau_2\left({\bf S}\cdot{\bf\Pi}+{\bf\Pi}\cdot{\bf S}
 - {\textstyle\frac{2}{d}}tr({\bf\Pi}\cdot{\bf S}){\bf 1}\right)= {\bf\Pi}^{eq}
\end{eqnarray}
where
\begin{eqnarray*}
 {\bf\Pi}^{eq} &=& -2\mu{\bf S} - \left(\lambda{\bf\nabla}{\bf\cdot}
        {\bf u}\right){\bf 1} + {\bf\Pi}_2 + {\bf\Pi}_3, \\
   {\bf S} &=& {\textstyle\frac{1}{2}}\left({\bf L} + {\bf L}^T\right) -
                {\textstyle\frac{1}{d}}\left({\bf\nabla}{\bf\cdot}{\bf u}\right) {\bf 1}, \\
   {\bf L} &=& ({\bf\nabla\bf u})^T = \left(\Frac{\partial u_i}{\partial x_j}\right), \\
   \lambda &=& (\zeta - \frac{2}{d}\mu), \\
 \tau_1 &=& 0.3211\left(\frac{\mu}{p}\right), \\
 \tau_2 &=&  0.58775\left(\frac{\mu}{p}\right).  
\end{eqnarray*}
Here $\tau_1$ and $\tau_2$ are relaxation times, $d$ the dimensionality 
of the system, $\bf L$ is the velocity gradient, $\bf S$ the deviatoric 
part of the rate of strain tensor, $\mu$ the shear viscosity, 
$\zeta$ the bulk viscosity and $\bf 1$ the identity tensor; 
${\bf\Pi}_2$ and ${\bf\Pi}_3$ are higher order terms, 
explicitly written down in Jin \& Slemrod$^{51}$.
Note that both relaxation times are proportional 
to the ratio of the shear viscosity and the pressure,
and hence proportional to the mean free time. 
In the limits of $\tau_1$, $\tau_2 \to 0$ and 
${\bf\Pi}_2$, ${\bf\Pi}_3 \to 0$, we recover
the standard Newtonian model for the stress tensor.

Neglecting the higher-order terms, an expression for the first normal stress
difference can be obtained for the steady uniform shear flow:
\begin{equation}
{\mathcal N}_1 = \left[\frac{4\tau_1}{2(1+\tau_1^2)
                             +\tau_2(1+\tau_1-\tau_2)}\right]
                \left(\frac{\mu}{p}\right) \;\;>\;\; 0.
\end{equation}
This quantity is always positive, as in our simulation results for dilute flows,
if the two relaxation times are of the same order.

It is important to note that the above  evolution equation 
does not satisfy the {\it principle of material frame indifference} (MFI)
which states that the constitutive laws should be {\it invariant}
under rigid-rotation$^{7,8}$.
The scalar field $\phi$, the vector field $\bf v$ 
and the tensor field ${\bf \Pi}$ are called 
{\it frame-indifferent} or {\it objective} if the following relations hold
for all $t$:
\begin{eqnarray}
   \phi'({\bf x}',t') &=& \phi({\bf x},t), \\
  {\bf v}'({\bf x}',t')&=&  {\bf Q}(t){\bf v}({\bf x},t), \\ 
  {\bf \Pi}'({\bf x}',t') &=& {\bf Q}(t){\bf \Pi}({\bf x},t){\bf Q}(t)^T, 
\end{eqnarray}
where $\phi$, $\bf v$ \& ${\bf \Pi}$ and $\phi'$, $\bf v'$ \& ${\bf \Pi}'$ 
are defined in two different frames
${\mathcal F}$ and ${\mathcal F}'$, respectively,
and ${\bf Q}(t)$ is a proper orthogonal tensor. 
Here  ${\mathcal F}'\in {\mathcal E}({\mathcal F})$,
with ${\mathcal E}({\mathcal F})$
denoting the set of all frames obtainable from a given frame
${\mathcal F}$ by observer transformations.
That the stress-tensor in a granular gas is not a frame-independent 
quantity (as in the hard-sphere gas$^{54,55}$)
is well-known. 
Since the kinetic component of the first normal stress
difference remains positive at all densities, the kinetic stress tensor 
can be modelled using a non-objective equation as discussed above.
For the collisional stress tensor, one can postulate a similar
evolution equation as in Eqn. (11), but $\tau_1$ must be
multiplied by a factor which must change sign at the critical density.
However, this equation would remain frame-dependent even in the dense limit.
Thus, if one has to recover the Boltzmann-limit of relaxation-type equations,
a frame-indifferent approach does not appear to work.

A question now arises as to the possibility of modelling
normal stress differences using the standard frame-indifferent relaxation type models.
The simplest way is to use either the {\it lower-convected}
or the {\it upper-convected} equations
for the pressure deviator:
\begin{eqnarray}
 {\bf\Pi} + \alpha(\nu,e)\tau\left(\Frac{{\rm D}{\bf\Pi}}{{\rm D}t} 
   + \left\{ {\bf L}^T \cdot{\bf\Pi} 
   + {\bf\Pi} \cdot {\bf L} 
   - {\textstyle\frac{2}{d}}{\rm tr}({\bf\Pi}\cdot{\bf L}){\bf 1} \right\} \right)
  &=& -2\mu{\bf S} - \left(\lambda{\bf\nabla}{\bf\cdot}{\bf u}\right){\bf 1}, 
\label{eq:rstress22}\\
 {\bf\Pi} + \alpha(\nu,e)\tau\left(\Frac{{\rm D} {\bf\Pi}}{{\rm D}t} 
   - \left\{ {\bf L}\cdot{\bf\Pi} 
   + {\bf\Pi}\cdot{\bf L}^T 
    - {\textstyle\frac{2}{d}}{\rm tr}({\bf\Pi} \cdot {\bf L}){\bf 1} \right\} \right)
  &=& -2\mu{\bf S} - \left(\lambda{\bf\nabla}{\bf\cdot}{\bf u}\right){\bf 1}, 
\label{eq:rstress21}
\end{eqnarray}
respectively.
Here $\alpha(\nu,e)$ is an empirical constant, dependent on both
the density and restitution coefficient.
For both cases, the first normal stress difference is
\begin{equation}
  {\mathcal N}_1 =  -\frac{2\tau\alpha}{1+\tau^2\alpha^2}\left(\frac{\mu}{p}\right).
\label{eq:rnsd_uc}
\end{equation}
Clearly, if $\alpha(\nu,e)$ is obtained from simulation, its
sign-reversal would also correspond to the sign-reversal of ${\mathcal N}_1$.
Comparing the Jin-Slemrod equation with its corresponding
frame-indifferent analog (lower-convected model), we  conclude that
the loss of frame-indifference shows up as a {\it sign-change} of the first
normal stress difference.
It would be interesting to investigate whether
one could relax the Sela-Goldhirsch equations$^{4}$ using
a frame-indifferent approach without violating the entropy inequality$^{51}$.

Similarly, one could postulate evolution equations
using other objective derivatives.
In this regard, the co-rotational Jeffrey's model seems to be the 
ideal choice:
\begin{equation}
{\bf\Pi} + \tau_1(\nu,e)\Frac{{\mathcal D}{\bf\Pi}}{{\mathcal D}t} =
   -2\mu\left({\bf S} + \frac{\lambda}{2\mu}\left({\bf\nabla}{\bf\cdot}{\bf u}\right){\bf 1}
        + \tau_2(\nu,e)\Frac{{\mathcal D}{\bf S}}{{\mathcal D}t} \right)
\end{equation}
with ${\mathcal D}/{{\mathcal D}t}$ being the Jaumann derivative$^7$.
The corresponding first normal stress difference is
\begin{equation}
  {\mathcal N}_1 =  -\frac{2(\tau_1-\tau_2)}{1+\tau_1^2}\left(\frac{\mu}{p}\right),
\label{eq:rnsd_jef}
\end{equation}
which is positive/negative depending on whether $\tau_1$
is less/greater than  $\tau_2$.  
Thus, the frame-indifferent
relaxation models are able to predict positive and negative
first normal stress differences.
(For the steady homogeneous shear flow, one can also 
model positive/negative normal stress differences by postulating 
a general orthonormal basis, generated by the nilpotent basis tensors,
which satisfies the objectivity requirement; for related
issues, the reader is referred to Goddard$^8$.)

\section{Summary and Conclusion}
\label{sec:conclusion}

We have probed the non-Newtonian behaviour and the incipient 
crystalline-phase in a sheared, monodisperse, two-dimensional 
granular fluid. The standard event-driven technique is used
to simulate a box of hard-disks under homogeneous shear deformation.
The information about the stress tensor is obtained
by decomposing it in the standard way:
${\bf P}=p{\bf 1} + {\bf\Pi}$, where $p$ is
the pressure and $\bf\Pi$ the pressure deviator.
The non-Newtonian behaviour is quantified in terms
of the first  normal stress difference:
${\mathcal N}_1=({\Pi}_{xx}-{\Pi}_{yy})/p$.
 
The granular fluid is non-Newtonian with a measurable 
first normal stress difference (${\mathcal N}_1$)  
which is {\it positive} (if the stress
is defined in the {\it compressive} sense) in the dilute limit.
Interestingly, however, ${\mathcal N}_1$ {\it changes}
from  positive to negative at a critical density in the dense regime.
By decomposing ${\mathcal N}_1$ into the kinetic and
collisional contributions, ${\mathcal N}_1={\mathcal N}_1^k + {\mathcal N}_1^c$,
we found that while ${\mathcal N}_1^k$ 
is always positive and decays to zero in the dense limit,
${\mathcal N}_1^c$ has a {\it non-monotonic} variation
with density. In particular, ${\mathcal N}_1^c$ 
increases from zero in the dilute limit as $\nu$ increases, reaches a maximum
at some value of $\nu$ and then decreases, 
eventually becoming {\it negative} in the dense limit.
The density  at which ${\mathcal N}_1^c=0$ ($\nu\equiv\overline\nu$) 
depends crucially on the level of micro-scale dissipation;
in particular, $\overline\nu$ increases as the coefficient of restitution decreases.
We have constructed a phase-diagram in the ($\nu, e$)-plane 
by identifying the regions where ${\mathcal N}_1$ is positive/negative.

We have shown that the {\it origin} of the first normal stress difference, 
in the dense limit, is tied to shear-induced collisional anisotropies. 
The underlying mechanism is distinctly different from that 
is known for a dilute granular gas$^{4,16}$
where the anisotropy in the second moment of the fluctuation velcoity,
which is a Burnett-order effect,
gives rise  to normal stress differences.

At the micro-level, the particles undergo reorganization as the 
dense-limit is approached.
The signatures of microstructural-reorganization have been captured by
probing the collision-angle distribution, $C(\theta)$,
which is {\it anisotropic} due to the presence of the mean shear.
In particular, we have found that the topology of the
anisotropic-structure of $C(\theta)$ changes, with collisions
occurring at certain preferred angles on the upstream-faces of 
the colliding pairs.  The {\it sign-reversal} of ${\mathcal N}_1$
is  correlated with a preferred value of the
average collision angle, $\theta_{av}=\pi/4 \pm \pi/2$,
averaged over the {\it upstream}-faces of the colliding particles.

The time evolution of the sheared granular fluid 
leads to {\it crystallization} in the dense limit,
signalling the {\it coexistence} of fluid and solid.
The particles are arranged in a triangular-packing inside
the crystal, and it moves as a {\it layer} in the streamwise direction. 
The formation of such layered-crystalline structure depends 
crucially on the system size and the dissipation level:
the larger the system-size and the weaker the dissipation, 
the more susceptible the system is to crystallize. 
This appears to be related to a long-wave instability$^{56}$
of the elastic hard-sphere fluids.

The present work clearly shows that the available
kinetic-theory-based rheological models for granular fluids are not adequate to 
predict the behaviour of the first normal stress difference in the dense limit.
Certain microstructural-features, like the preferred distribution of collisions
which eventually leads to crystal-formation, should be incorporated
into the theory.
At such high densities, many-body effects (both positional and velocity
correlations) are important$^{57,41}$ and
the appropriate kinetic description is the BBGKY-hierarchy$^{57}$. 
To incorporate the observed normal stress behaviour into
the framework of plausible constitutive models,
we showed that the standard frame-indifferent relaxation type
models can be used to model both positive and negative
first normal stress differences.
In this regard, the two-parameter Jeffrey's model$^{7}$
appears to be the ideal choice;
however, we are unable to recover the corresponding
Boltzmann limit which is known to be non-objective.
On the whole, we believe that a lot remains to be done
for a better understanding of the dense-phase rheology
of granular fluids even in the hard-sphere limit.

\section{Acknowledgement}

M.A. acknowledges the financial support by the Alexander von Humboldt Foundation
and S.L. acknowledges the support of the
Deutsche Forschungsgemeinschaft. 
M.A. is grateful to Hans J. Herrmann for discussions and hospitality at ICA1,
and also acknowledges discussions with Joe D. Goddard on
certain aspects of normal stress behaviour in granular fluids.
We thank John F. Brady for directing us to relevant references
on normal stresses in suspensions.
We also thank three anonymous referees for their constructive comments.

\begin{appendix}

\section{Shear-induced anisotropy and the collision angle distribution}
\label{sec:theory}

Here we are interested in the shear-induced anisotropy of 
the collision angle distribution of an inelastic hard-disk fluid.
Following Savage \& Jeffrey$^{36}$,
an expression for the collision angle distribution $C(\theta)$ is derived,
which is compared with the simulation data in IIB.2.

To calculate the probablility of collisions at a specific angle $\theta$,
we focus on Fig. 13  with two particles colliding at ${\bf r}$.
Note that $\theta$ is measured anticlockwise from the positive $x$-axis.
For collisions to occur in a time $\delta t$, the center of particle $1$
must lie inside the volume $\sigma\delta{\bf k}({\bf q\cdot k})\delta t$,
where ${\bf q}={\bf c}_1-{\bf c}_2$ is the relative velocity of the colliding pair.
Thus the expected total number of collisions (per unit time and unit area) 
with the line of centres ${\bf k}$ lying 
between ${\bf k} - \delta{\bf k}/2$ and ${\bf k} + \delta{\bf k}/2$
is given by
\begin{equation}
 \int \sigma  f^{(2)}({\bf c}_1,{\bf r}_1,{\bf c}_2,{\bf r}_2)
({\bf q\cdot k})({\bf k\cdot n})d{\bf k}
   d{\bf c}_1 d{\bf c}_2,
\end{equation}
where $f^{(2)}(\cdot)$ is the  two-particle distribution function which is defined so that
$f^{(2)}({\bf c}_1,{\bf r}_1,{\bf c}_2,{\bf r}_2)$ $d{\bf c}_1d{\bf c}_2 d{\bf r}_1d{\bf r}_2$
is the number of pairs of particles such that the particle $i$ is located
in an area element $d{\bf r}_1$ about ${\bf r}_1$
with its velocity in the interval $d{\bf c}_1$ about ${\bf c}_1$
while particle $j$ is located
in an area element $d{\bf r}_2$ about ${\bf r}_2$
with its velocity in the interval $d{\bf c}_2$ about ${\bf c}_2$.
To progress further, we have to invoke the assumption of molecular chaos
and hence the expected number of collisions is
\begin{equation}
 \int \sigma g({\bf r}_1,{\bf r}_2) f^{(1)}({\bf c}_1,{\bf r}_1;{\bf u}({\bf r}_1))
  f^{(1)}({\bf c}_2,{\bf r}_2;{\bf u}({\bf r}_2))({\bf q\cdot k})({\bf k\cdot n})d{\bf k}
   d{\bf c}_1 d{\bf c}_2,
\end{equation}
where $g({\bf r}_1,{\bf r}_2)$ is the pair-distribution function.
For the steady uniform shear flow,  $g({\bf r}_1,{\bf r}_2)$
is calculated from the relation$^{35,36,58}$:
\begin{equation}
 g({\bf r}_1,{\bf r}_2) = \frac{2g_c(\nu)}{n^2}\int_{{\bf q}\cdot{\bf k}>0} 
  f^{(1)}({\bf c}_1,{\bf r}_1;{\bf u}({\bf r}_1))f^{(1)}({\bf c}_2,{\bf r}_2;{\bf u}({\bf r}_2)) d{\bf c}_1d{\bf c}_2,
\label{eqn_pcorr1}
\end{equation}
where $g_c(\nu)$ is the {\it contact} value of the pair-distribution function and
${\bf q}\cdot{\bf k}>0$ implies that the integration
be carried out for impending collisions. 

As a first approximation, the single particle velocity distribution function
$ f^{(1)}({\bf c}_1,{\bf r}_1;{\bf u}({\bf r}_1))$ is assumed to have the
Maxwellian-form:
\begin{equation} 
 f^{(1)}({\bf c}_1,{\bf r}_1;{\bf u}({\bf r}_1)) = \left(\frac{nm}{\pi k_B T}\right)
   \exp\left[-\frac{m\left({\bf c}_1-{\bf u}({\bf r}_1)\right)^2}{2k_BT}\right],
\end{equation}
where $T$ is the granular temperature (fluctuating kinetic energy) 
and $k_B$  the usual Boltzmann constant.
Now transforming the particle velocities (${\bf c}_1$, ${\bf c}_2$)
to their center-of-mass and relative velocities, 
equation (\ref{eqn_pcorr1}) can be integrated
to arrive at the following explicit expression for 
the pair-distribution function$^{36}$:
\begin{equation}
 g({\bf r}_1,{\bf r}_2) = g_c(\nu) {\rm erfc}\left[ \frac{2{\bf k}\cdot
    {\bf u}({\bf r}_2)}{(2k_B T)^{1/2}}\right],
\end{equation}
where ${\rm erfc}(\cdot)$ is the complementary error function.
Using the above expression for the pair-distribution  function
and transforming in terms of polar coordinates ($r,\theta$),
the integral for the normalized collision angle distribution 
yields$^{37}$ 
\begin{equation}
 C(\theta) = {\mathcal A}(T)\left[\exp\left(-\frac{\sin^2\theta\cos^2\theta}{2T}\right) 
          - \frac{g(\theta)\sin\theta\cos\theta}{\sqrt{T}} \right]g(\theta),
\end{equation}
where $g(\theta)$ is the angular pair-distribution function
given by
\begin{equation}
  g(\theta) \equiv \frac{g({\bf r}_1,{\bf r}_2)}{g_c(\nu)}
    = {\rm erfc}\left[ \frac{\sin\theta\cos\theta}{\sqrt{2T}} \right]
\end{equation}
and ${\mathcal A}(T)$ is a normalization constant.

For the uniform shear flow, an expression for the granular temperature,
can be obtained from the energy balance
equation, by equating the energy production due to shear-work
with the energy loss due to collisions:
\begin{eqnarray}
 \mu\left(\frac{du}{dy}\right)^2 &=& {\mathcal D} \nonumber \\
\Rightarrow T &=& f_{\mu}(\nu)/f_{\mathcal D}(\nu,e)
\end{eqnarray}
where $\mu=\rho_p\sigma f_{\mu}(\nu)\sqrt{T}$ is the shear viscosity
and ${\mathcal D}=(\rho_p/\sigma)f_{\mathcal D}(\nu,e)T^{3/2}$ 
the collisional dissipation rate,
with
\begin{eqnarray*}
 f_{\mu}(\nu)&=&\frac{\sqrt{\pi}\nu}{8}\left[\frac{1}{\nu g_c(\nu)}+2+\nu g_c(\nu)\left(1+\frac{8}{\pi}\right)\right], \\
 f_{\mathcal D}(\nu,e) &=& \frac{4}{\sqrt{\pi}}(1-e^2)\nu^2g_c(\nu).
\end{eqnarray*}
Substituting this expression for $T$,
the normalized collision distribution function becomes
\begin{equation}
 C(\theta; \nu, e)={\mathcal A}(T)\left[\exp\left(-\frac{f_{\mathcal D}(\nu,e)\sin^2\theta\cos^2\theta}{2f_{\mu}(\nu)}\right) 
          - g(\theta)\sin\theta\cos\theta \sqrt{\frac{f_{\mathcal D}(\nu,e)}{f_{\mu}(\nu)}} \right]g(\theta)
\label{eqn_colang1}
\end{equation}
and the angular pair-distribution function
\begin{equation}
  g(\theta; \nu, e) 
    = {\rm erfc}\left[ \sin\theta\cos\theta\sqrt{\frac{f_{\mathcal D}(\nu,e)}{2f_{\mu}(\nu)}} \right].
\end{equation}
It is clear that the angular dependence of $C(\theta;\nu,e)$ is  modified 
by both the inelastic dissipation and the density.

\end{appendix}

\begin{figure}[htbp]
\begin{center}
\parbox{0.62\textwidth}{\epsfig{file=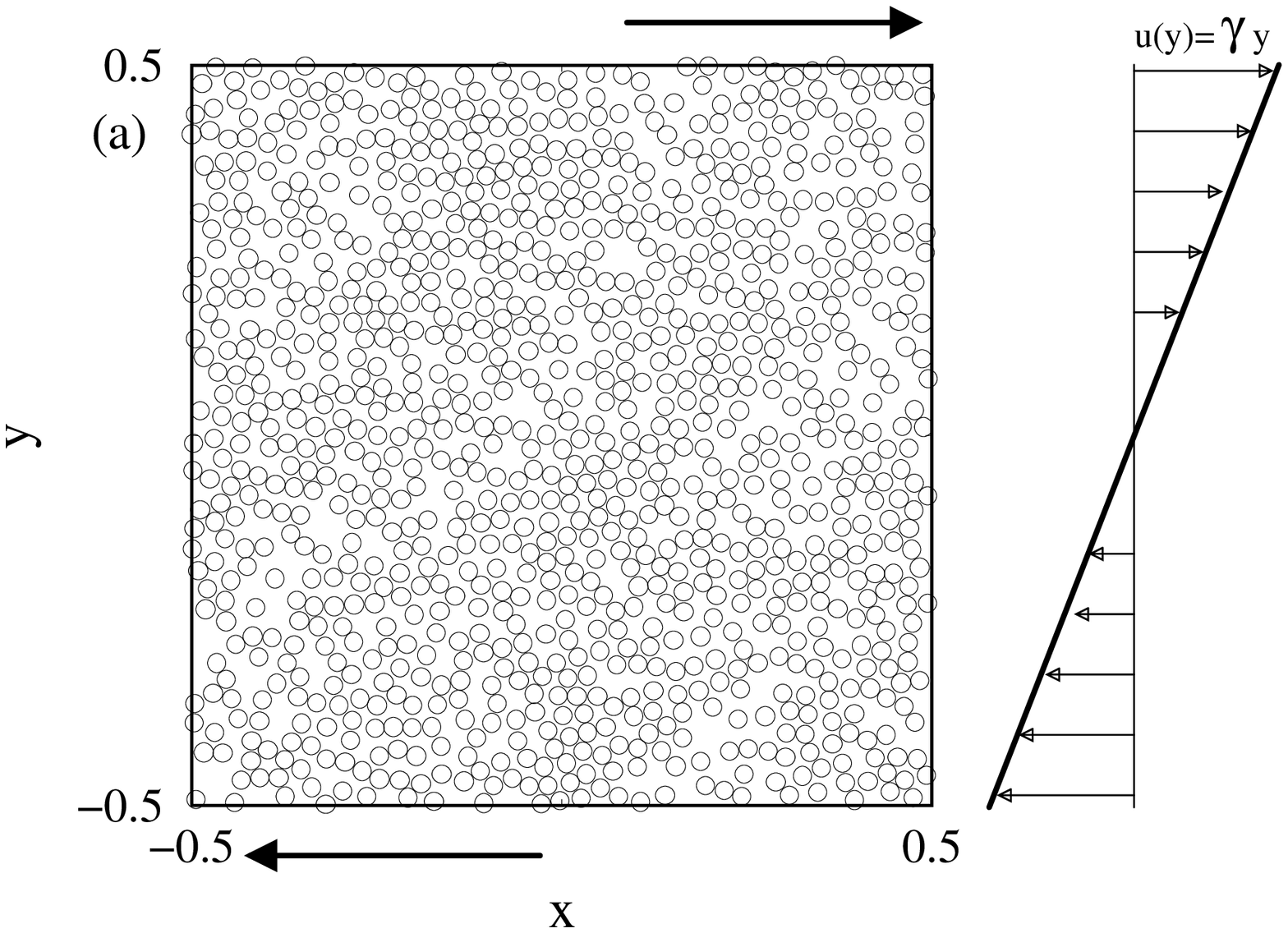,width=0.62\textwidth,angle=-00}} \\
\parbox{0.62\textwidth}{\epsfig{file=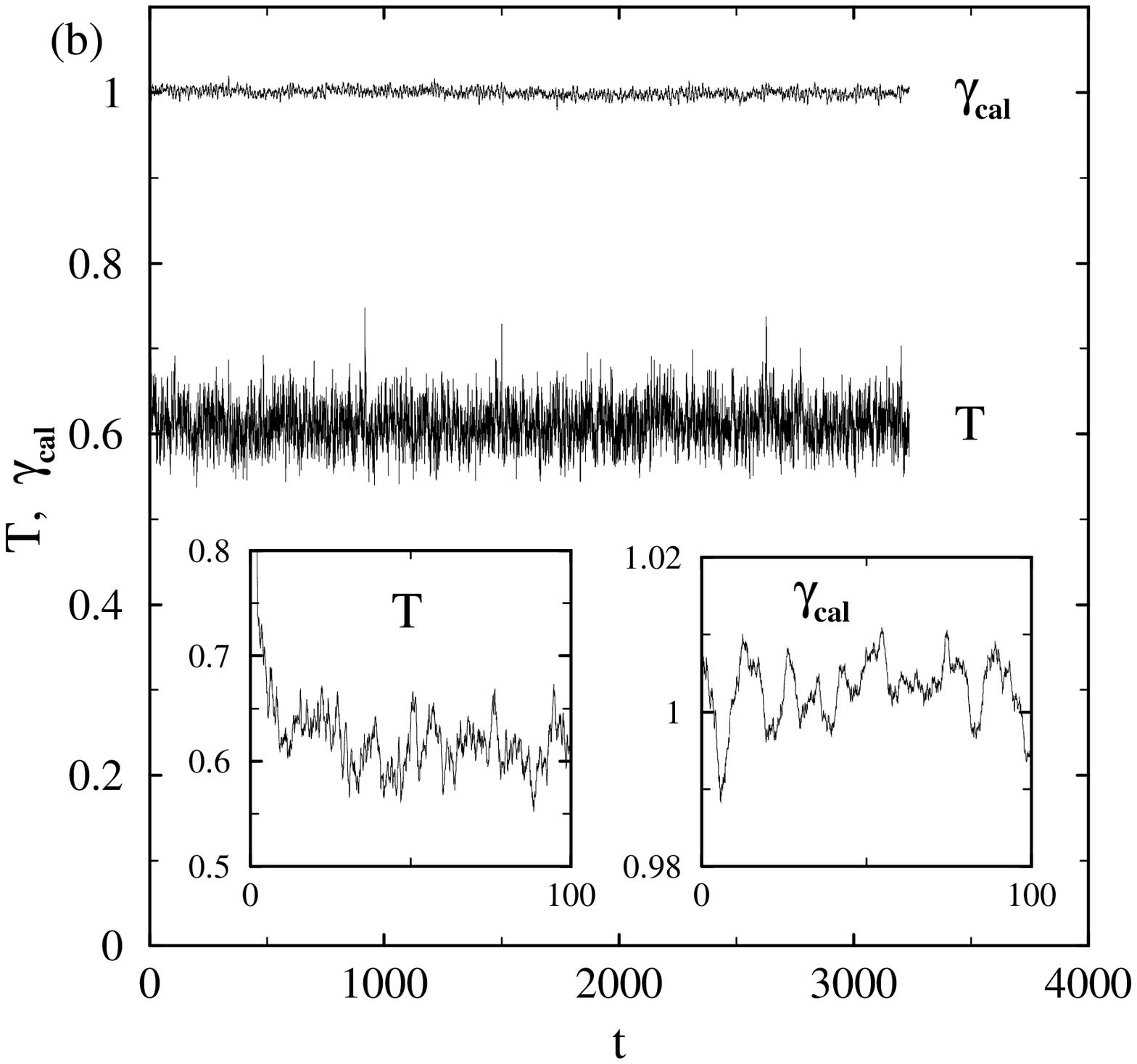,width=0.62\textwidth,angle=-00}}
\end{center}
\caption{
($a$) A snapshot of the sheared  granular system at
steady-state. The arrows indicate the displacement of the image boxes.
($b$) Variations of the granular energy $T$ and the calculated
shear rate $\gamma_{cal}$ with time. For an explanation of the
system and particle properties, see the text.
The parameters for both subplots are $\nu=0.5$,
$e=0.7$ and $N=1024$.
}
\label{fig:fig1}
\end{figure}

\newpage
\vspace*{-1.50cm}
\begin{figure}[htbp]
\begin{center}
\parbox{0.60\textwidth}{
 \epsfig{file=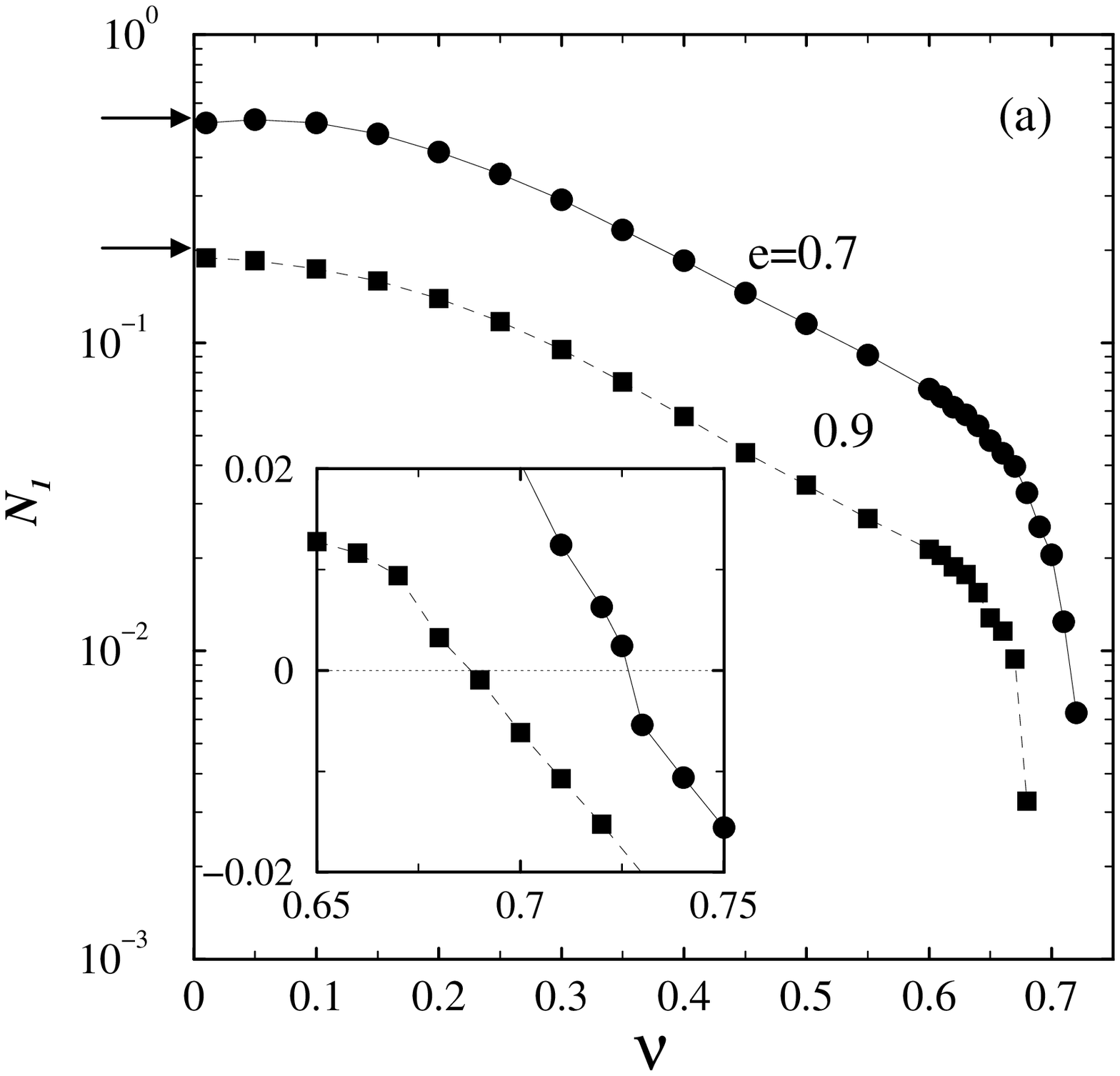,width=0.60\textwidth,angle=-00}} \\
\parbox{0.60\textwidth}{
 \epsfig{file=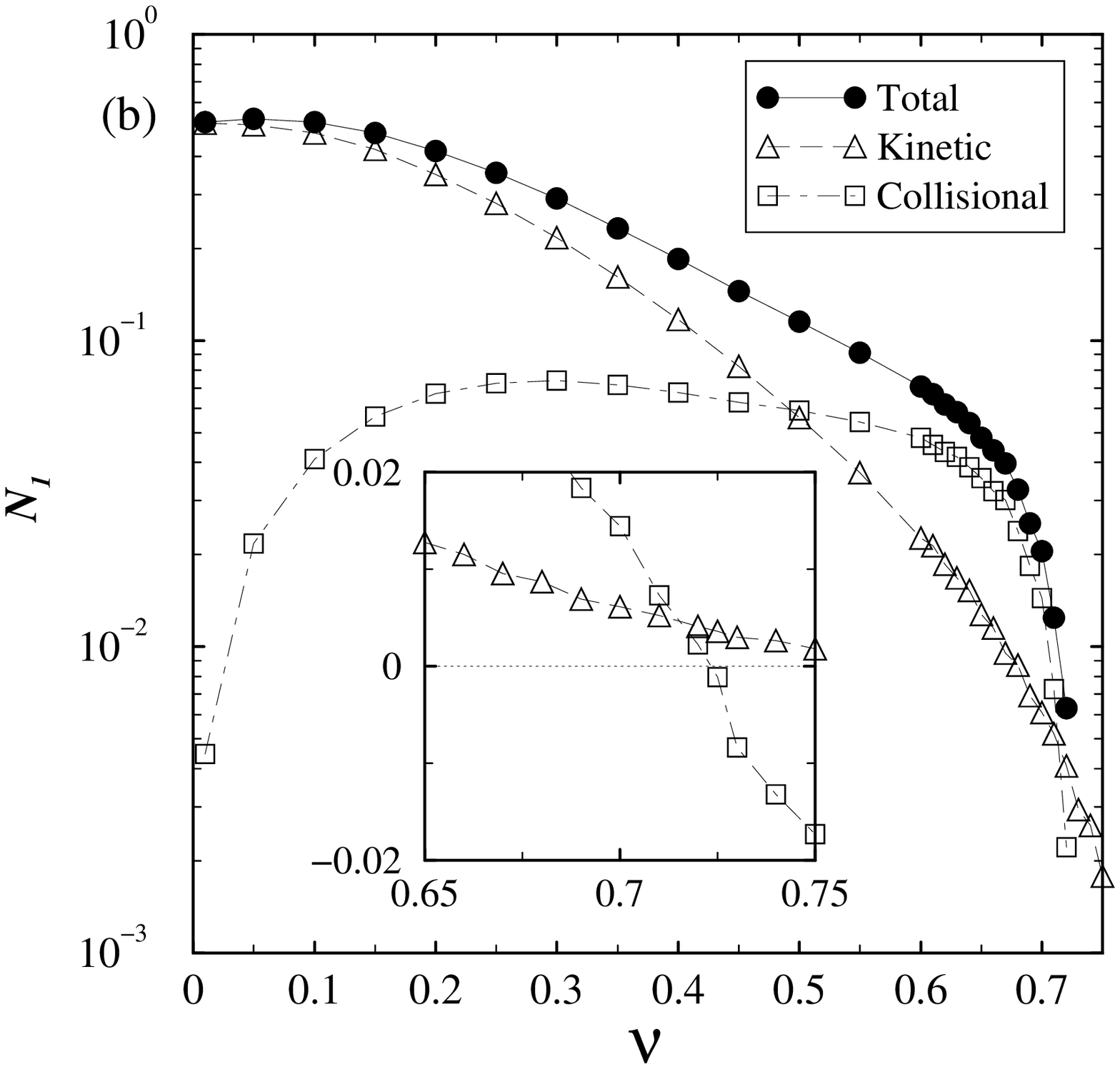,width=0.60\textwidth,angle=-00}}
\end{center}
\caption{
(a) Variation of the first normal stress difference ${\mathcal N}_1$
with the solid volume fraction.
The arrows on the
left ordinate indicate corresponding analytical values for a two-dimensional
granular gas.
(b) Variations of ${\mathcal N}_1$, ${\mathcal N}_1^k$ and ${\mathcal N}_1^c$
with $\nu$ at $e=0.7$.
In both subplots, the symbols represent the
simulation data and the lines are drawn to guide the eye.
}
\label{fig:fig2}
\end{figure}

\newpage
\begin{figure}[htbp]
\begin{center}
\parbox{0.75\textwidth}{
 \epsfig{file=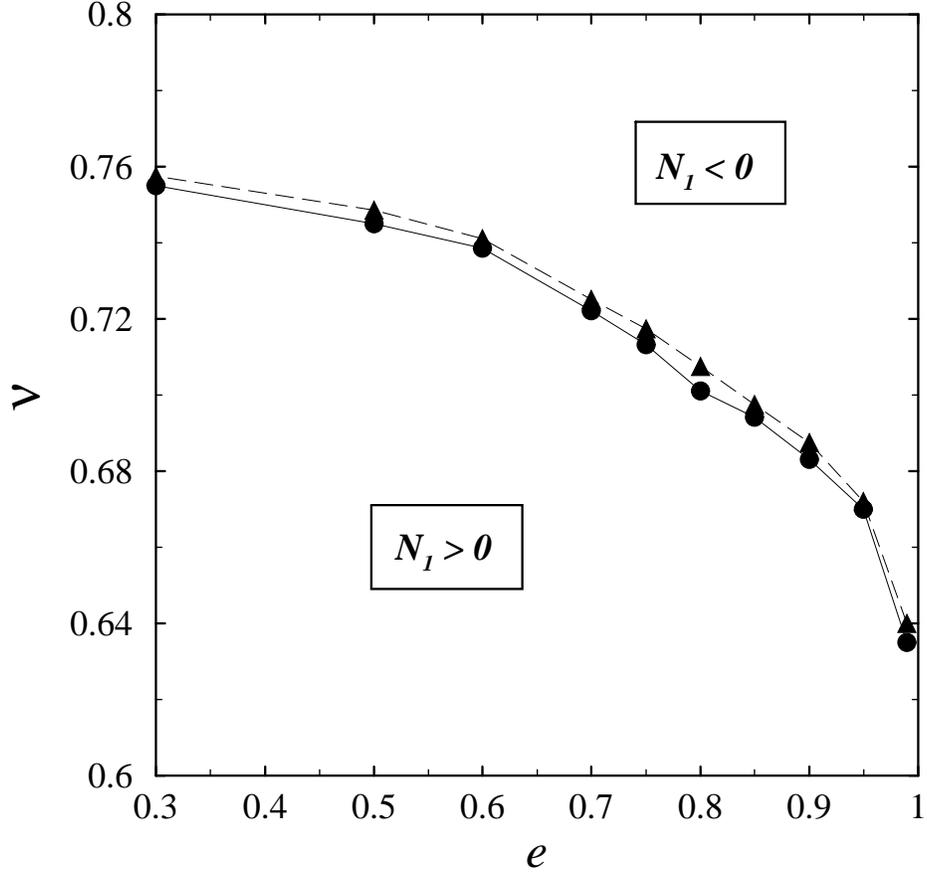,width=0.75\textwidth,angle=-00}} \\
\end{center}
\caption{
The phase diagram, delineating regions of positive and negative
first normal stress difference, in the ($\nu, e$)-plane. The filled
circles and triangles represent zeros of
${\mathcal N}_1^c$ and ${\mathcal N}_1$, respectively.
}
\label{fig:fig3}
\end{figure}

\newpage
\begin{figure}[htbp]
\begin{center}
\parbox{0.48\textwidth}{
 \epsfig{file=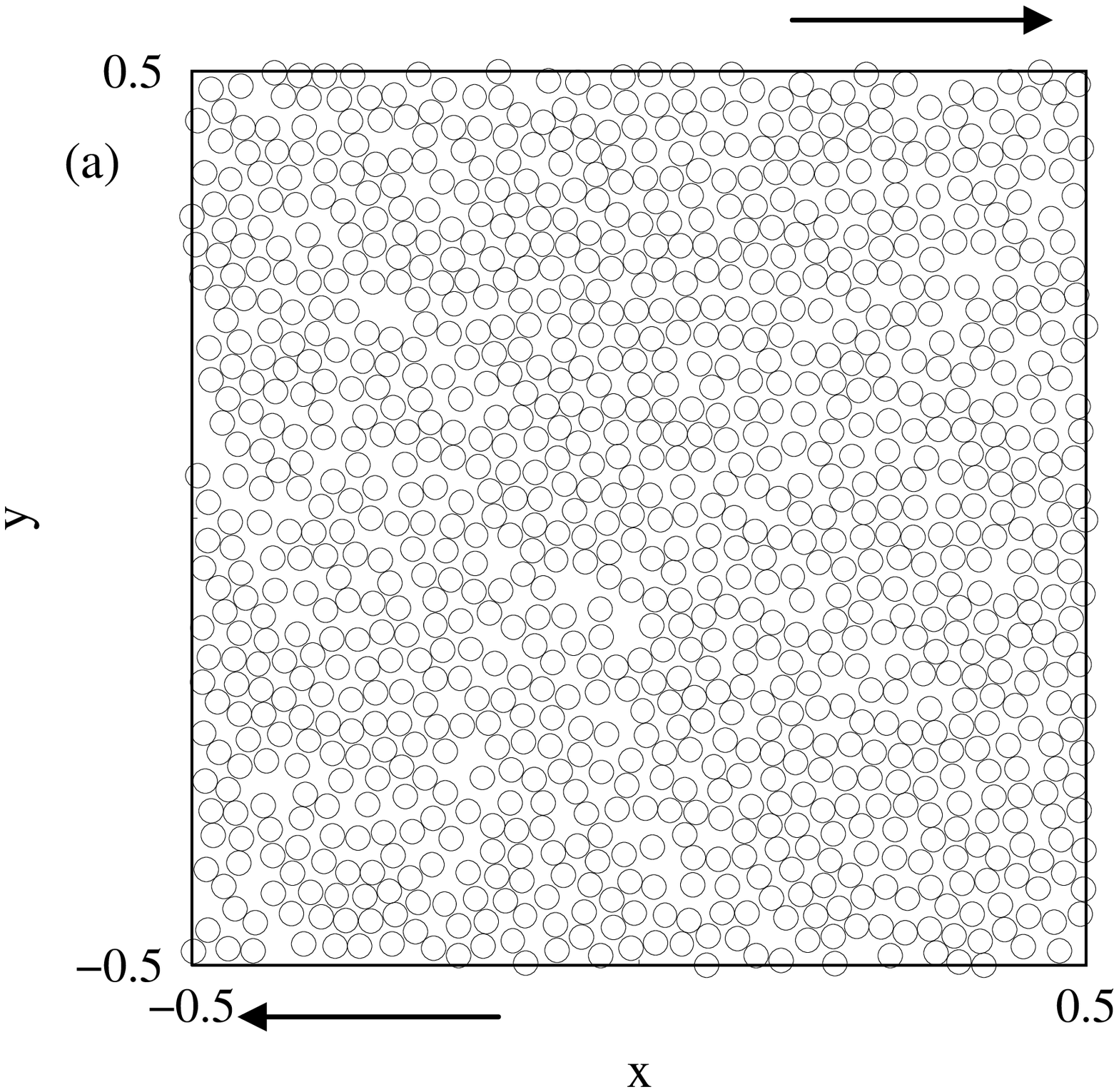,width=0.48\textwidth,angle=-00}}
\parbox{0.48\textwidth}{
 \epsfig{file=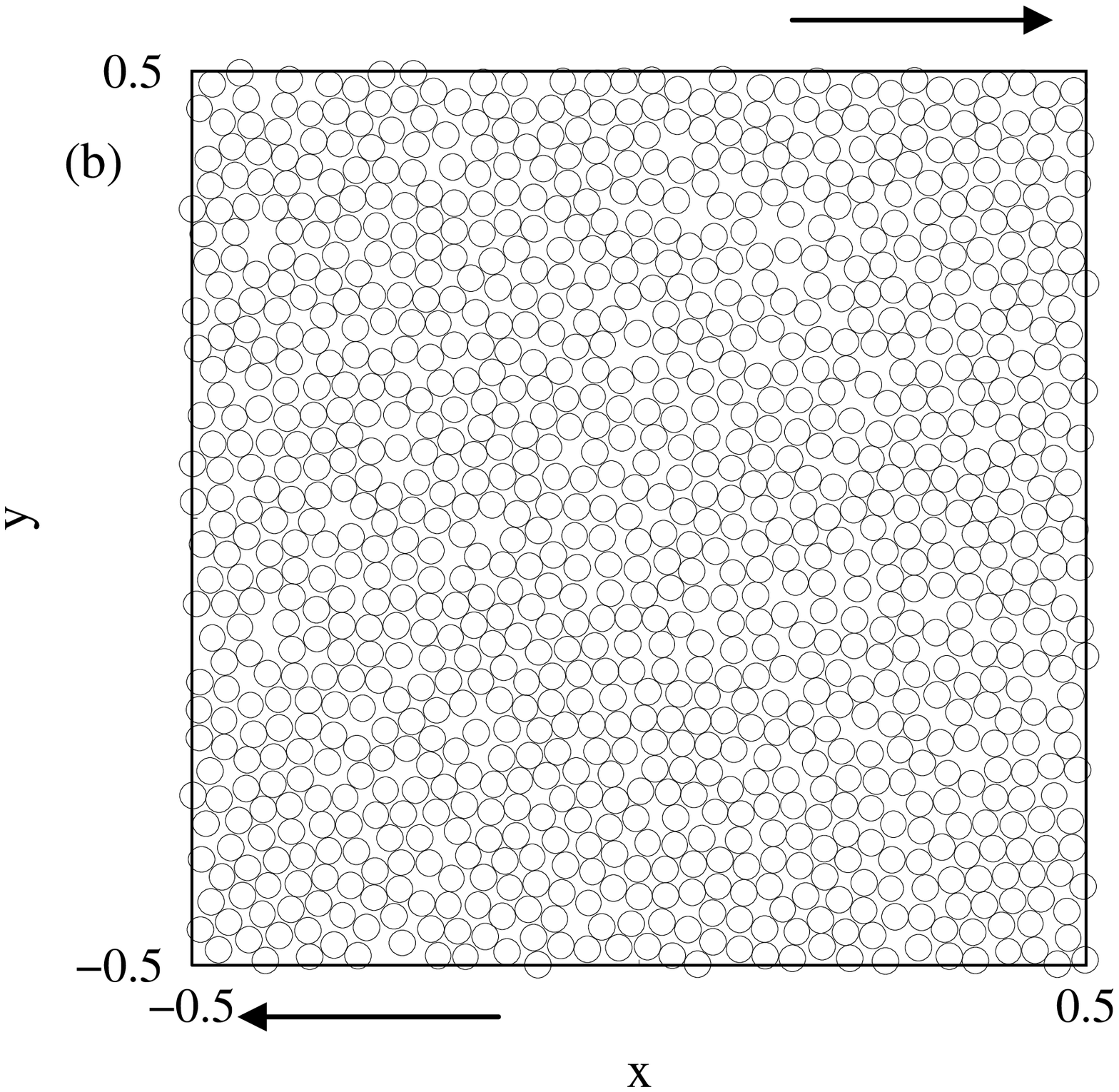,width=0.48\textwidth,angle=-00}} \\
\parbox{0.48\textwidth}{
 \epsfig{file=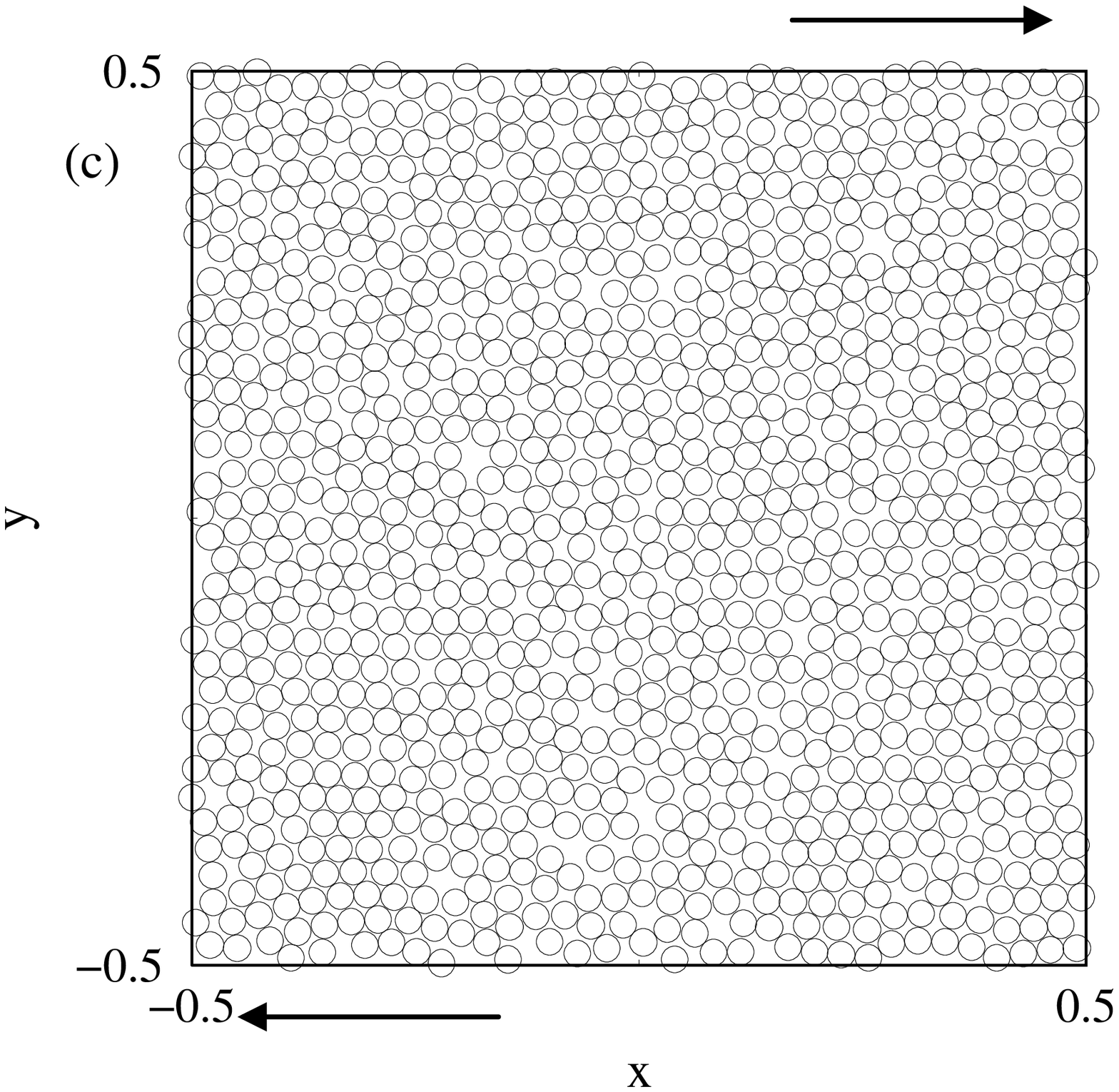,width=0.48\textwidth,angle=-00}}
\parbox{0.48\textwidth}{
 \epsfig{file=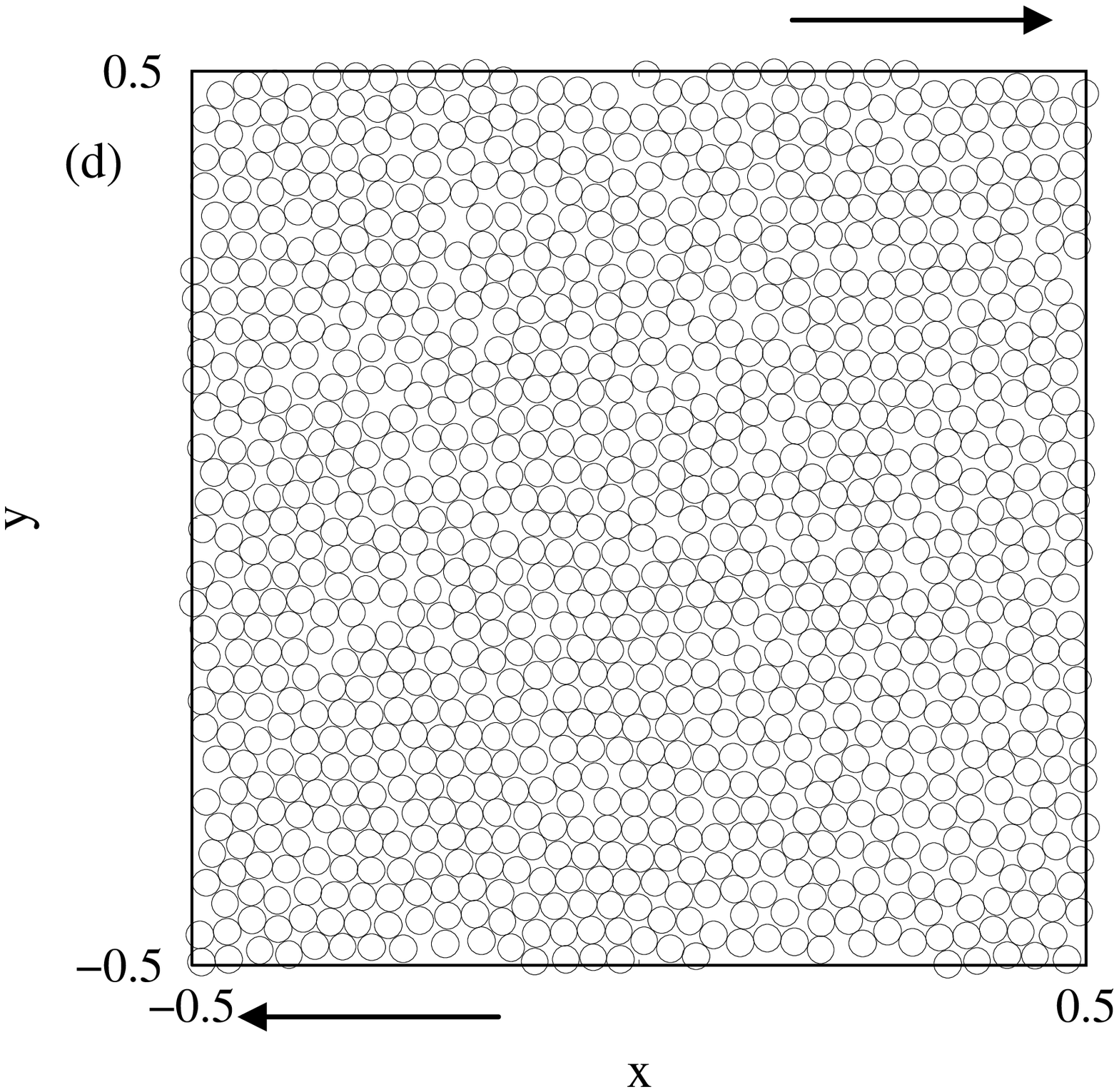,width=0.48\textwidth,angle=-00}}
\end{center}
\caption{
Snapshots of the sheared system with  $e=0.7$
in the dense limit for different densities: ($a$) $\nu=0.6$, ($b$) $\nu=0.7$;
($c$) $\nu=0.725$; ($d$) $\nu=0.75$.
}
\label{fig:fig4}
\end{figure}

\newpage
\begin{figure}[htbp]
\begin{center}
\parbox{0.42\textwidth}{
 \epsfig{file=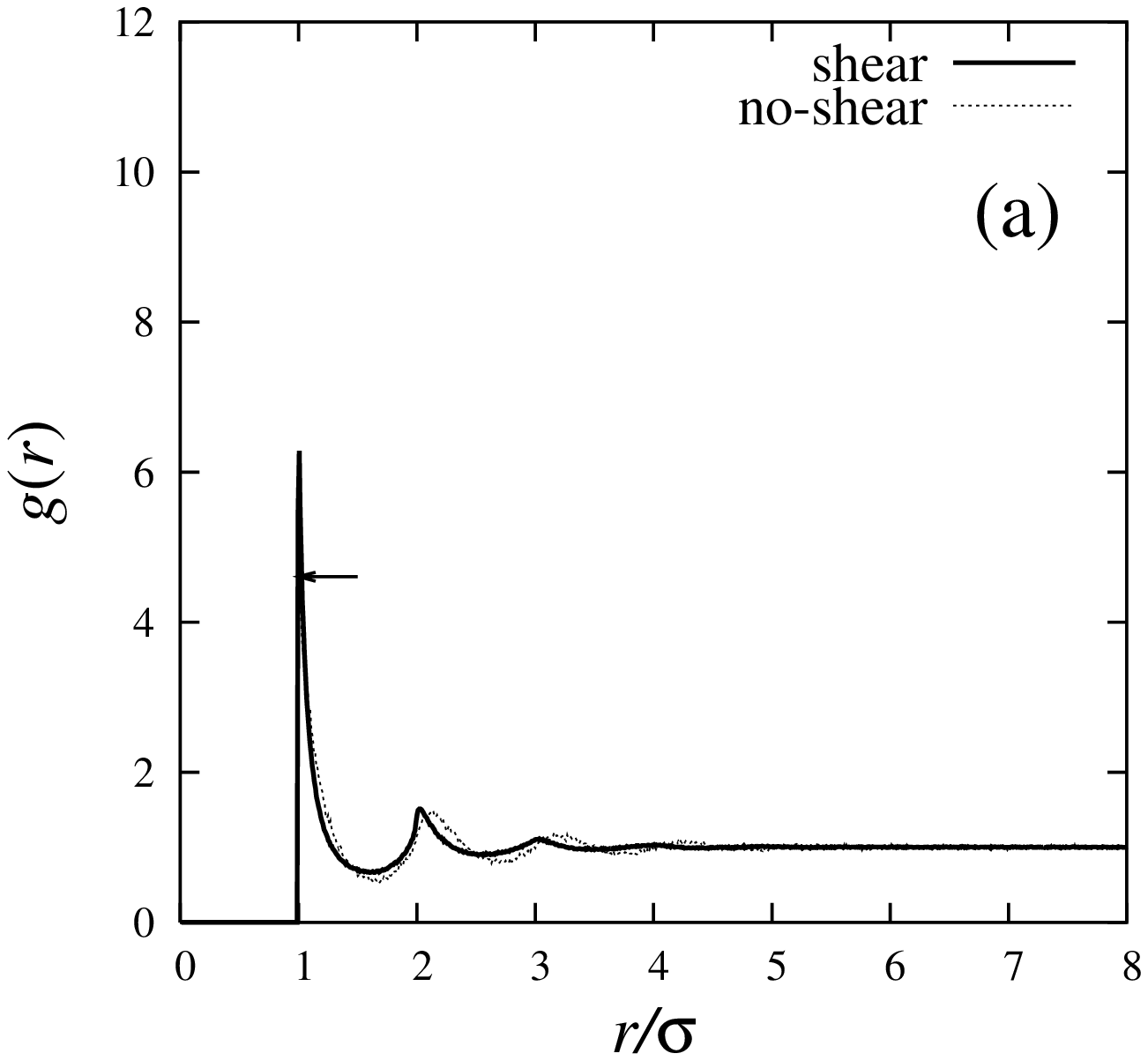,width=0.40\textwidth,angle=-90}}
\parbox{0.42\textwidth}{
 \epsfig{file=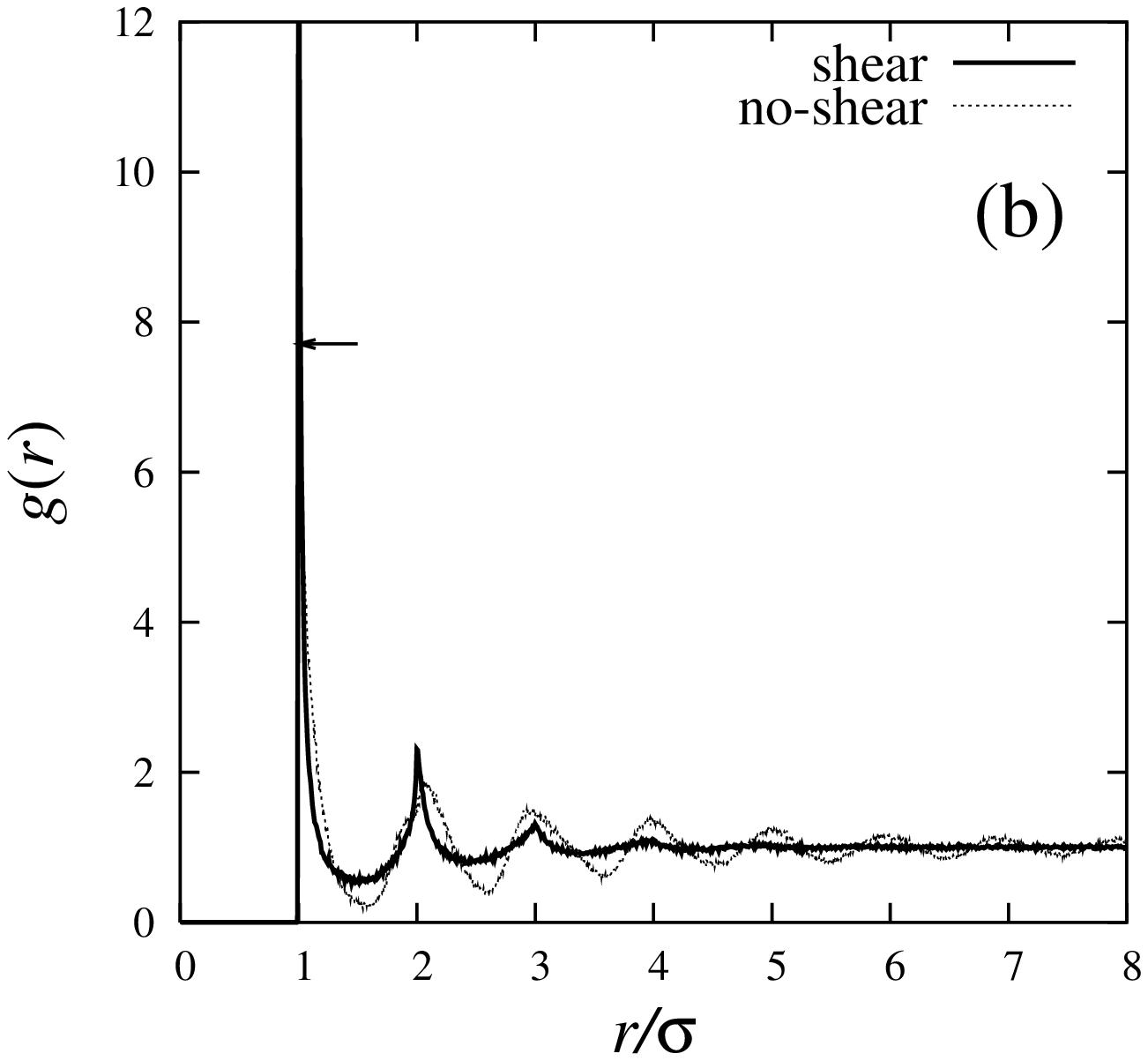,width=0.40\textwidth,angle=-90}} \\
\parbox{0.42\textwidth}{
 \epsfig{file=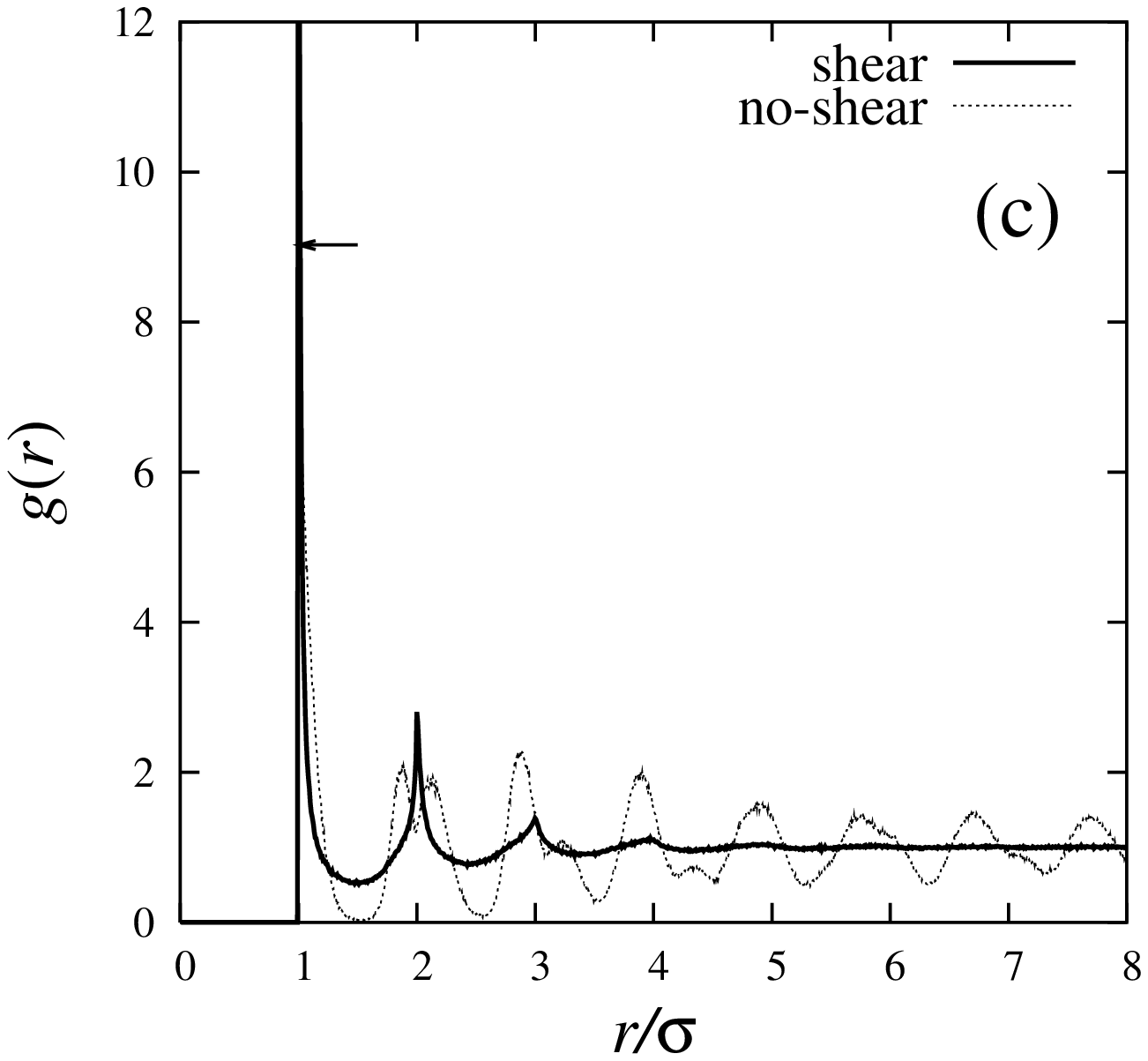,width=0.40\textwidth,angle=-90}}
\parbox{0.42\textwidth}{
 \epsfig{file=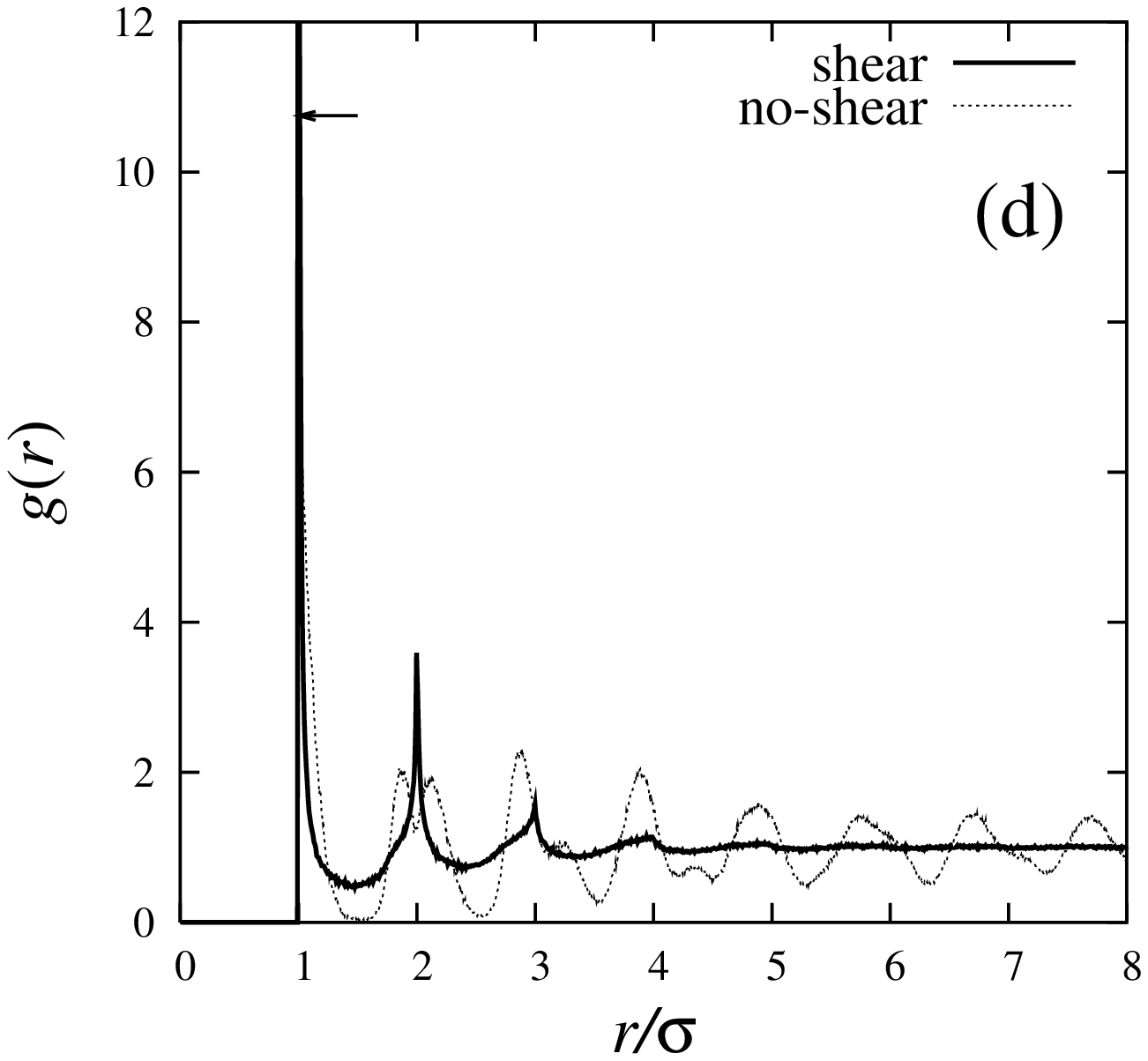,width=0.40\textwidth,angle=-90}}
\end{center}
\caption{
Radial distribution function $g(r)$ plotted against
the normalized distance $r/\sigma$ from sheared simulations with
dissipation $e=0.7$ (solid lines) and from homogeneous, non-sheared
situations with $\gamma=0$ and $e=1$ (dotted lines):
($a$) $\nu=0.6$, ($b$) $\nu=0.7$;
($c$) $\nu=0.725$; ($d$) $\nu=0.75$.
The arrows indicate the peak values of $g(r)$ at contact for the
homogeneous system.
}
\label{fig:fig6}
\end{figure}

\newpage
\begin{figure}[htbp]
\begin{center}
\parbox{0.75\textwidth}{
 \epsfig{file=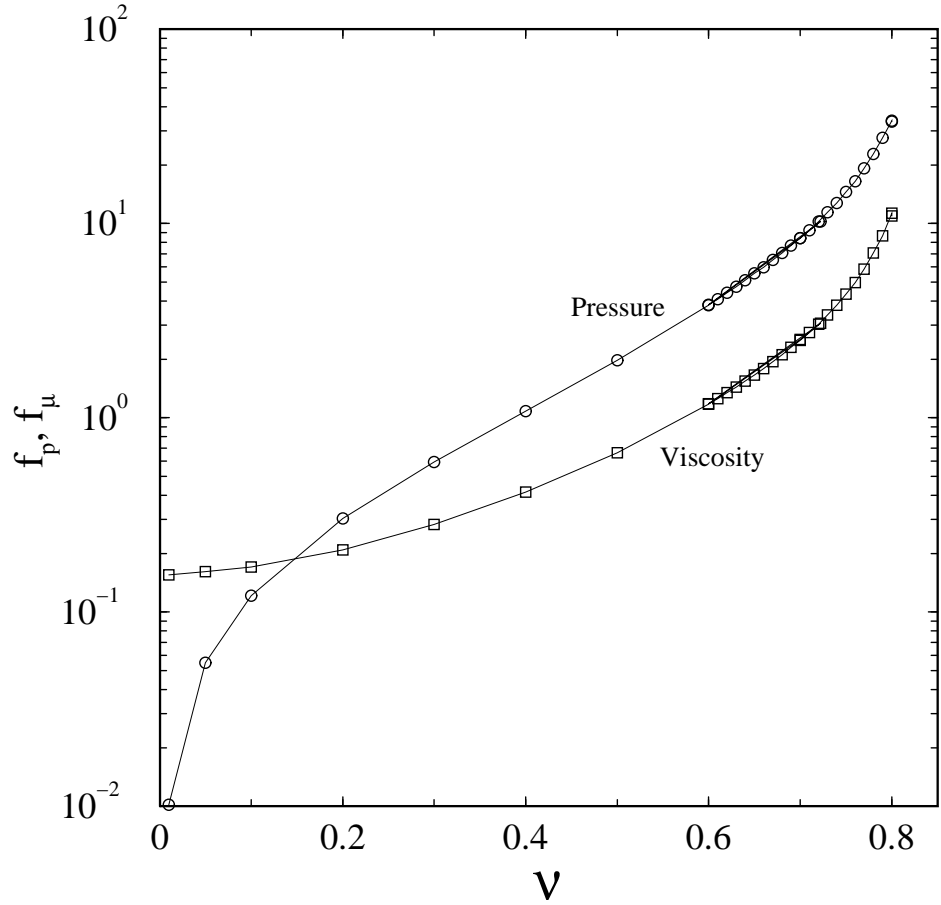,width=0.75\textwidth,angle=-00}} \\
\end{center}
\caption{
Variations of the pressure and viscosity functions with density at $e=0.7$:
$f_p=p/\rho T$ and $f_{\mu}=\mu/\rho\sigma\sqrt{T}$.
The lines are drawn to guide the eye.
}
\label{fig:fig7}
\end{figure}

\newpage
\begin{figure}[htbp]
\begin{center}
\parbox{0.75\textwidth}{
 \epsfig{file=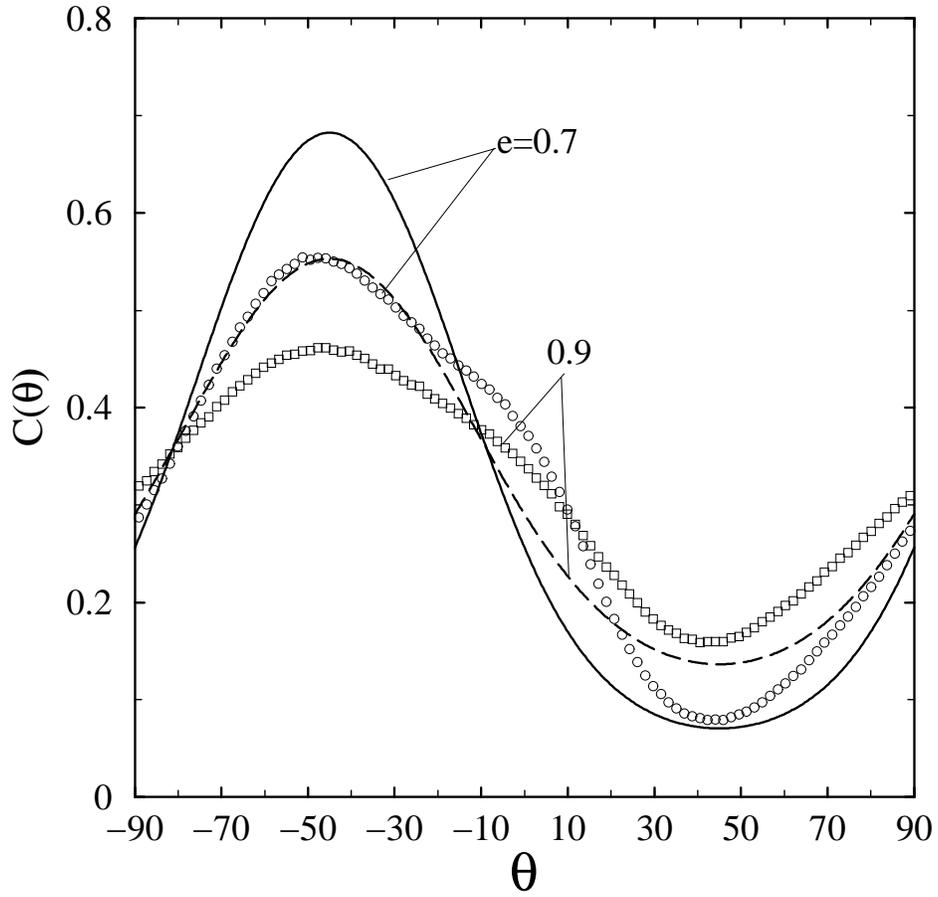,width=0.75\textwidth,angle=-00}} \\
\end{center}
\caption{
Distribution of collision angles $C(\theta)$ for different
coefficient of restitutions at $\nu=0.6$.
The symbols represent simulation data and the lines theoretical predictions.
}
\label{fig:fig8}
\end{figure}

\newpage
\begin{figure}[htbp]
\begin{center}
\parbox{0.48\textwidth}{
 \epsfig{file=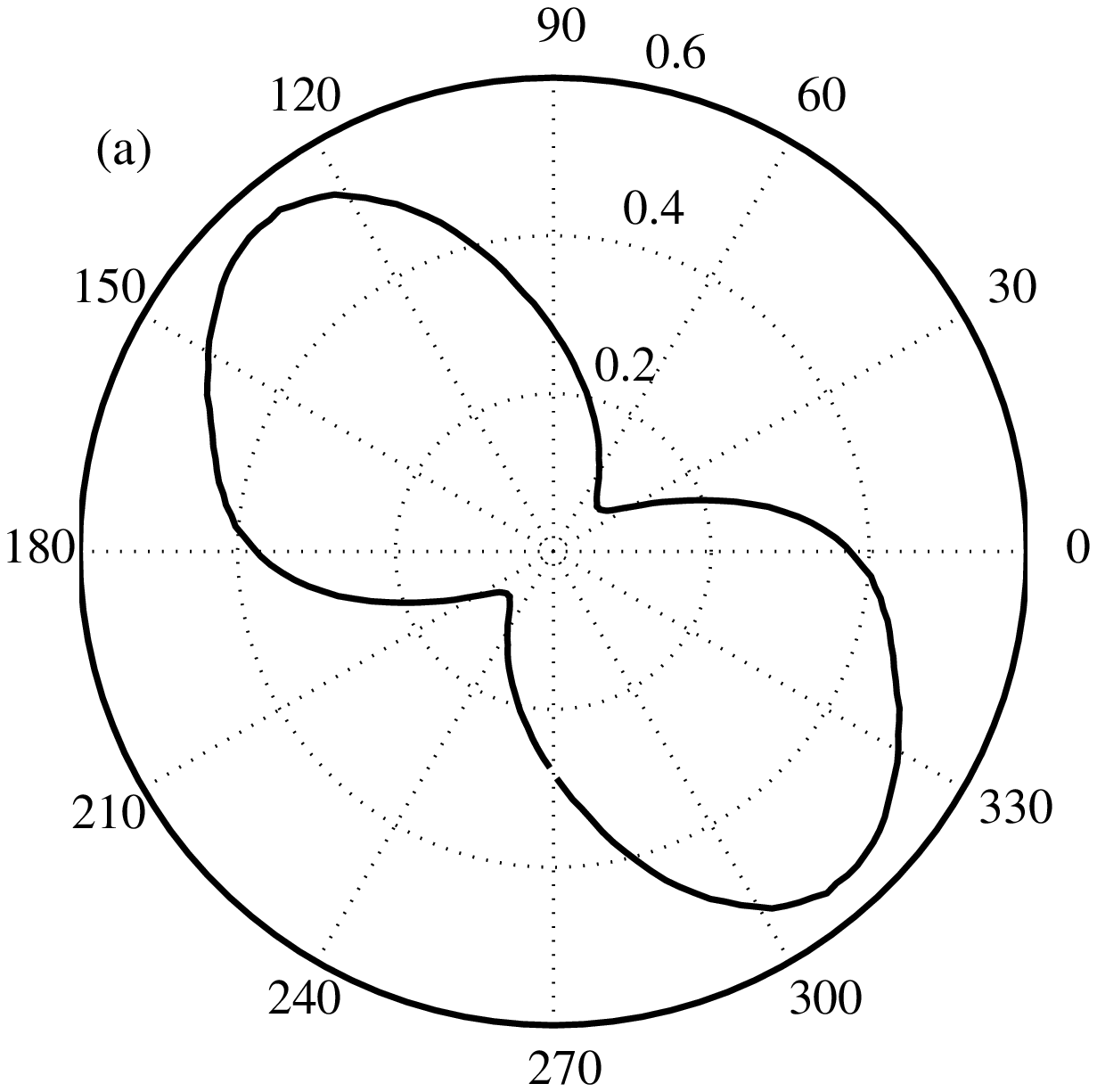,width=0.48\textwidth,angle=-00}}
\parbox{0.48\textwidth}{
 \epsfig{file=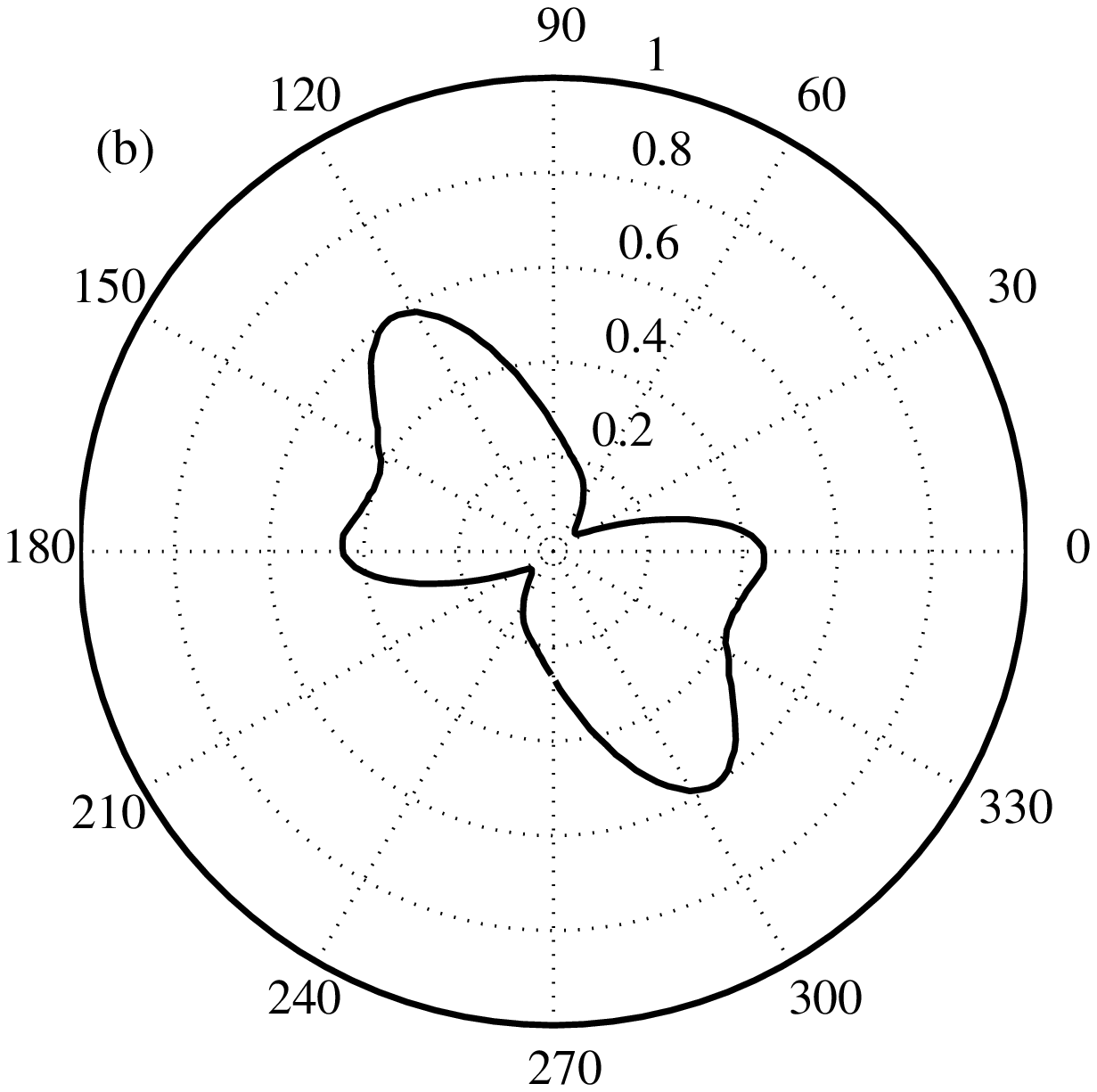,width=0.48\textwidth,angle=-00}} \\
\parbox{0.48\textwidth}{
 \epsfig{file=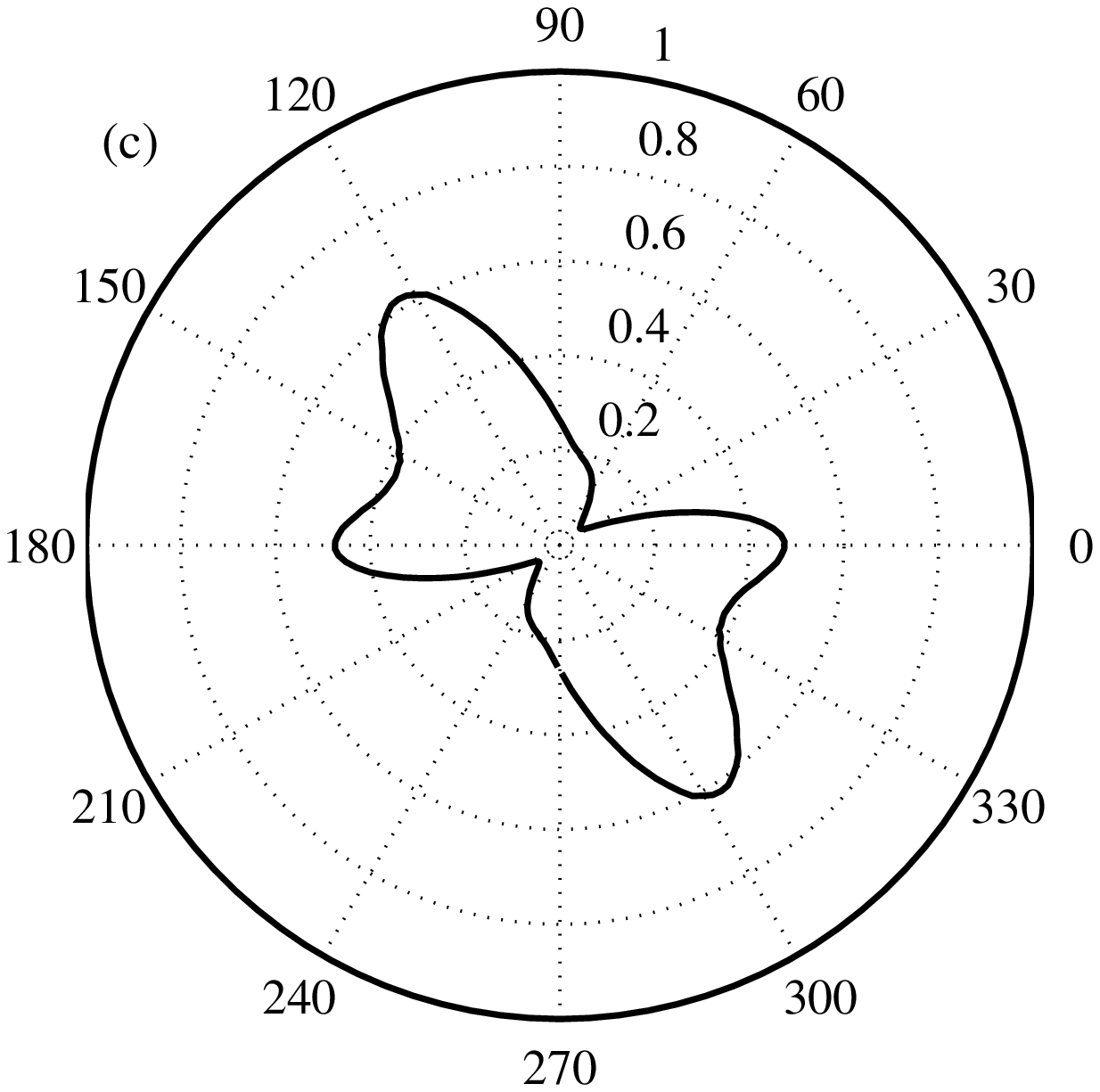,width=0.48\textwidth,angle=-00}}
\parbox{0.48\textwidth}{
 \epsfig{file=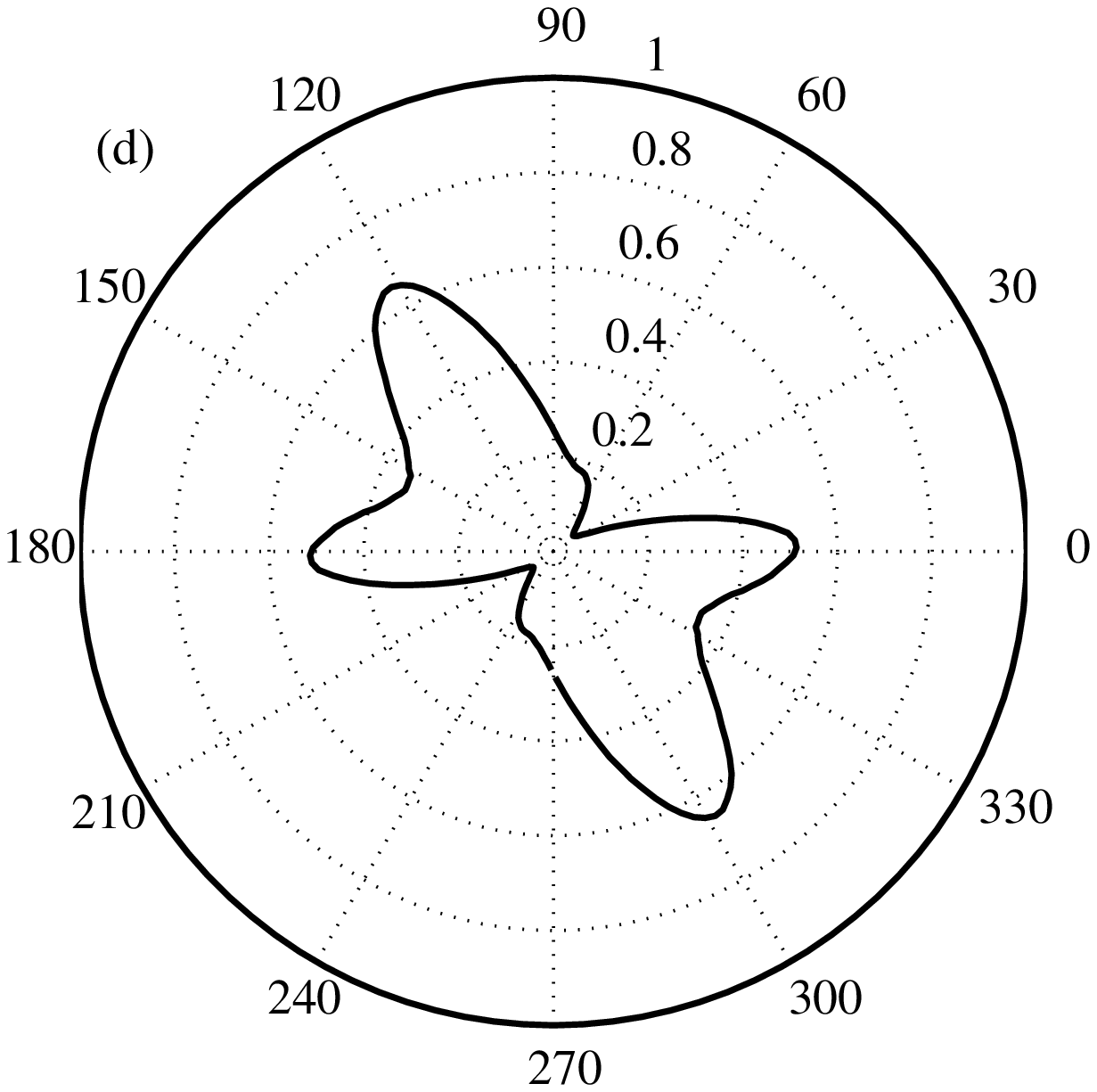,width=0.48\textwidth,angle=-00}}
\end{center}
\caption{
Polar plots of the collision angle distribution for different densities at  $e=0.7$:
($a$) $\nu=0.6$, ($b$) $\nu=0.7$;
($c$) $\nu=0.725$; ($d$) $\nu=0.75$.
}
\label{fig:fig9p}
\end{figure}

\newpage
\begin{figure}[htbp]
\begin{center}
\parbox{0.75\textwidth}{
 \epsfig{file=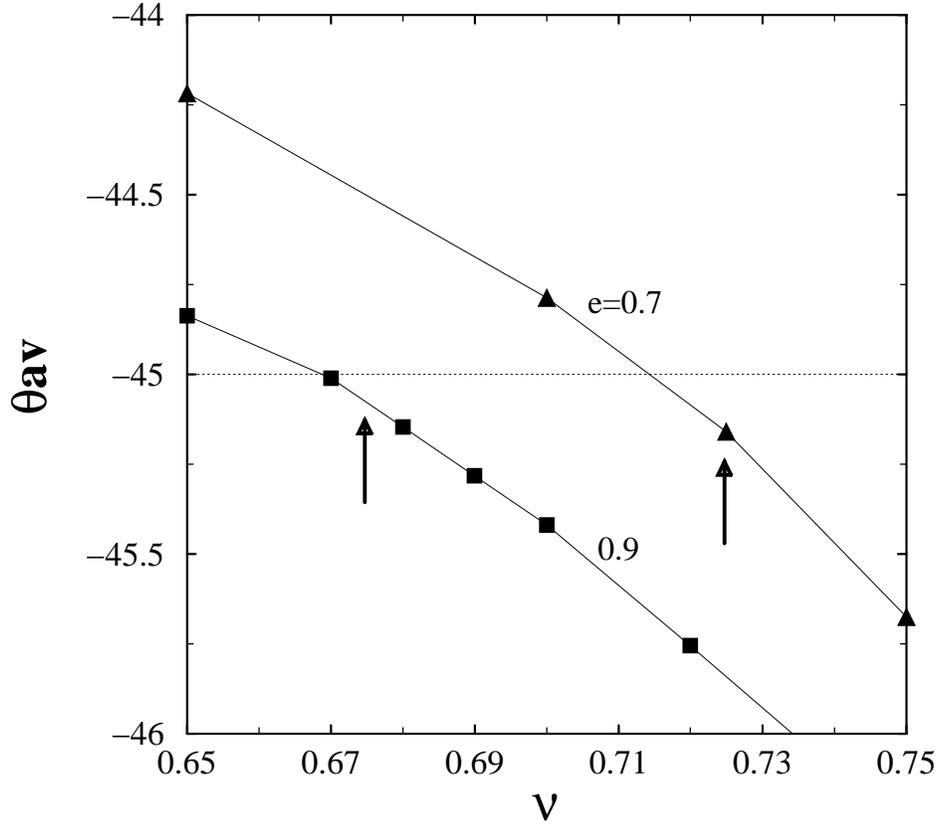,width=0.75\textwidth,angle=-00}} \\
\end{center}
\caption{
Variations of the average collision angle $\theta_{av}$ with density
for different restitution coefficients.
The arrows indicate densities where ${\mathcal N}_1\approx 0$.
The lines joining the data points are  to guide the eye.
}
\label{fig:fig10}
\end{figure}

\newpage
\begin{figure}[htbp]
\begin{center}
\parbox{0.48\textwidth}{\epsfig{file=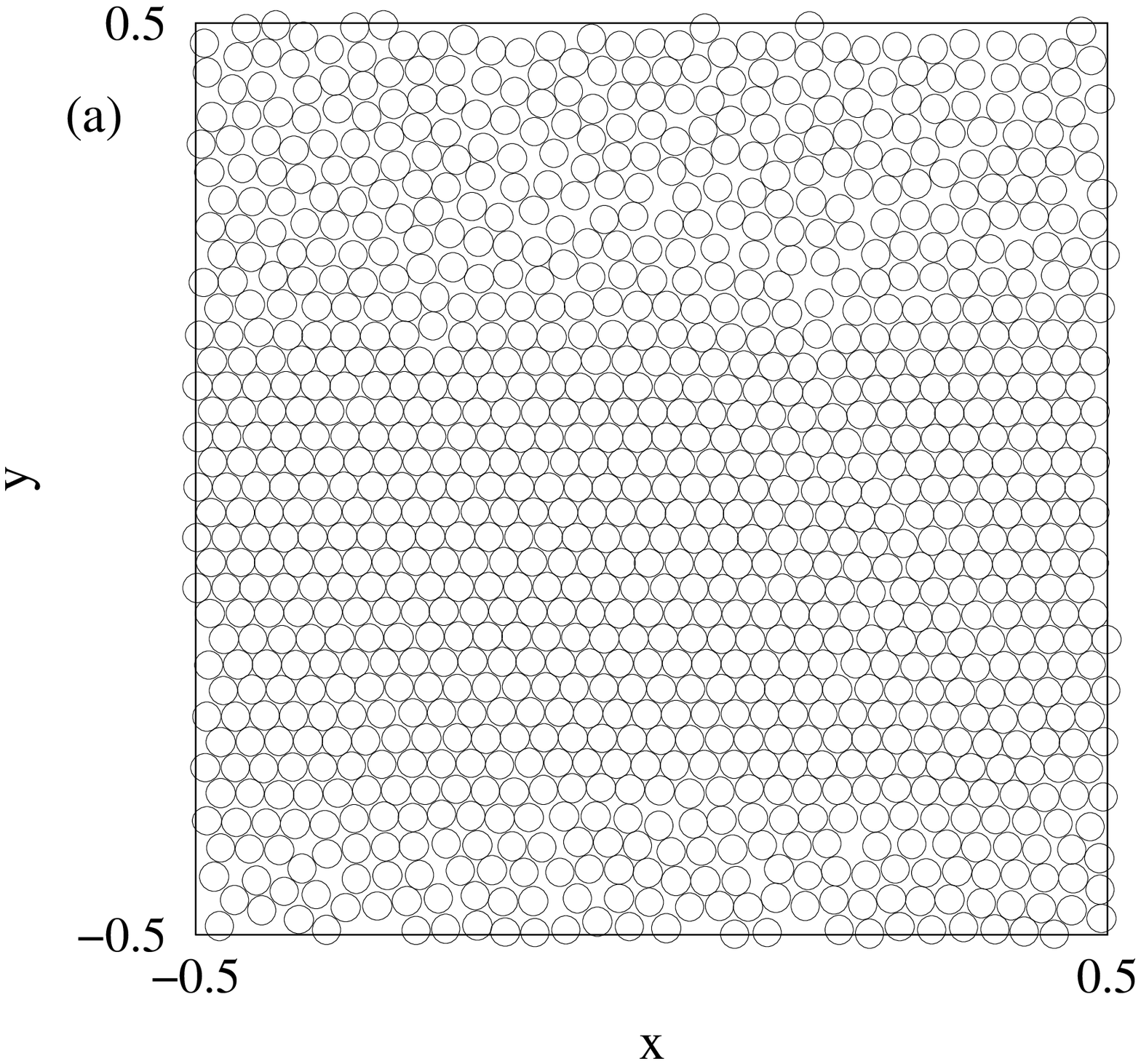,width=0.48\textwidth,angle=-00}}
\parbox{0.48\textwidth}{\epsfig{file=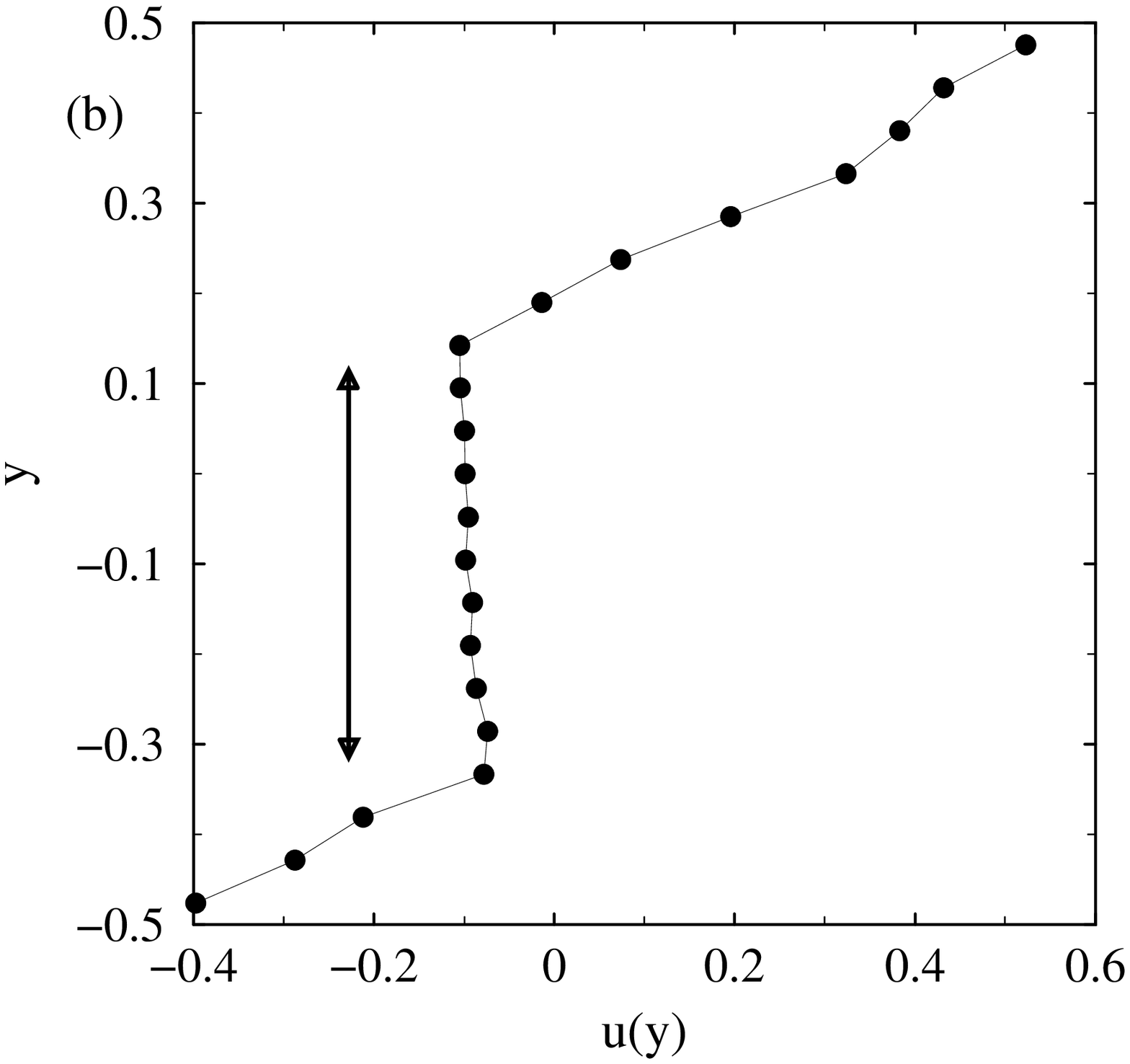,width=0.48\textwidth,angle=-00}}\\
\parbox{0.48\textwidth}{\epsfig{file=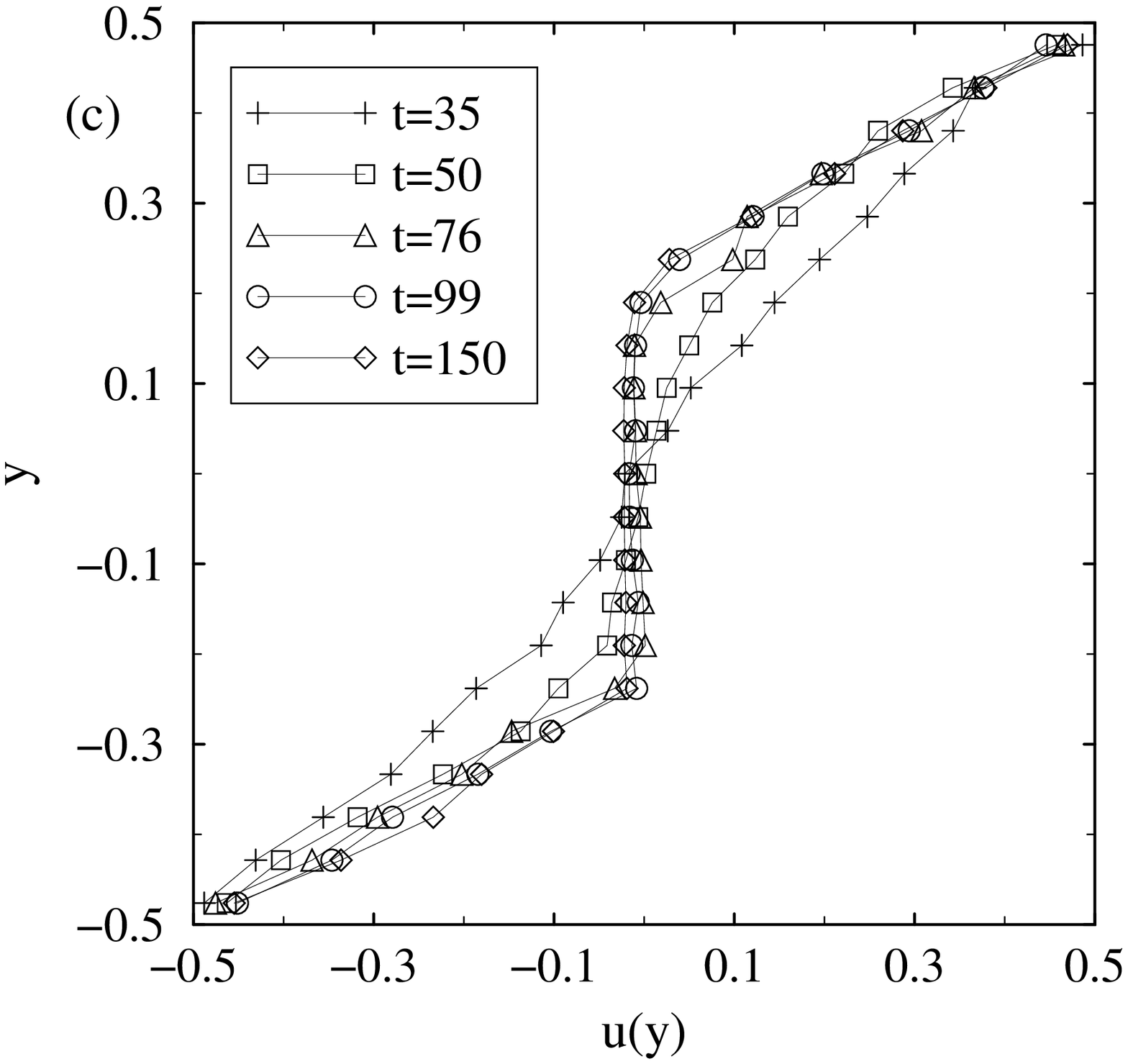,width=0.48\textwidth,angle=-00}}
\parbox{0.48\textwidth}{\epsfig{file=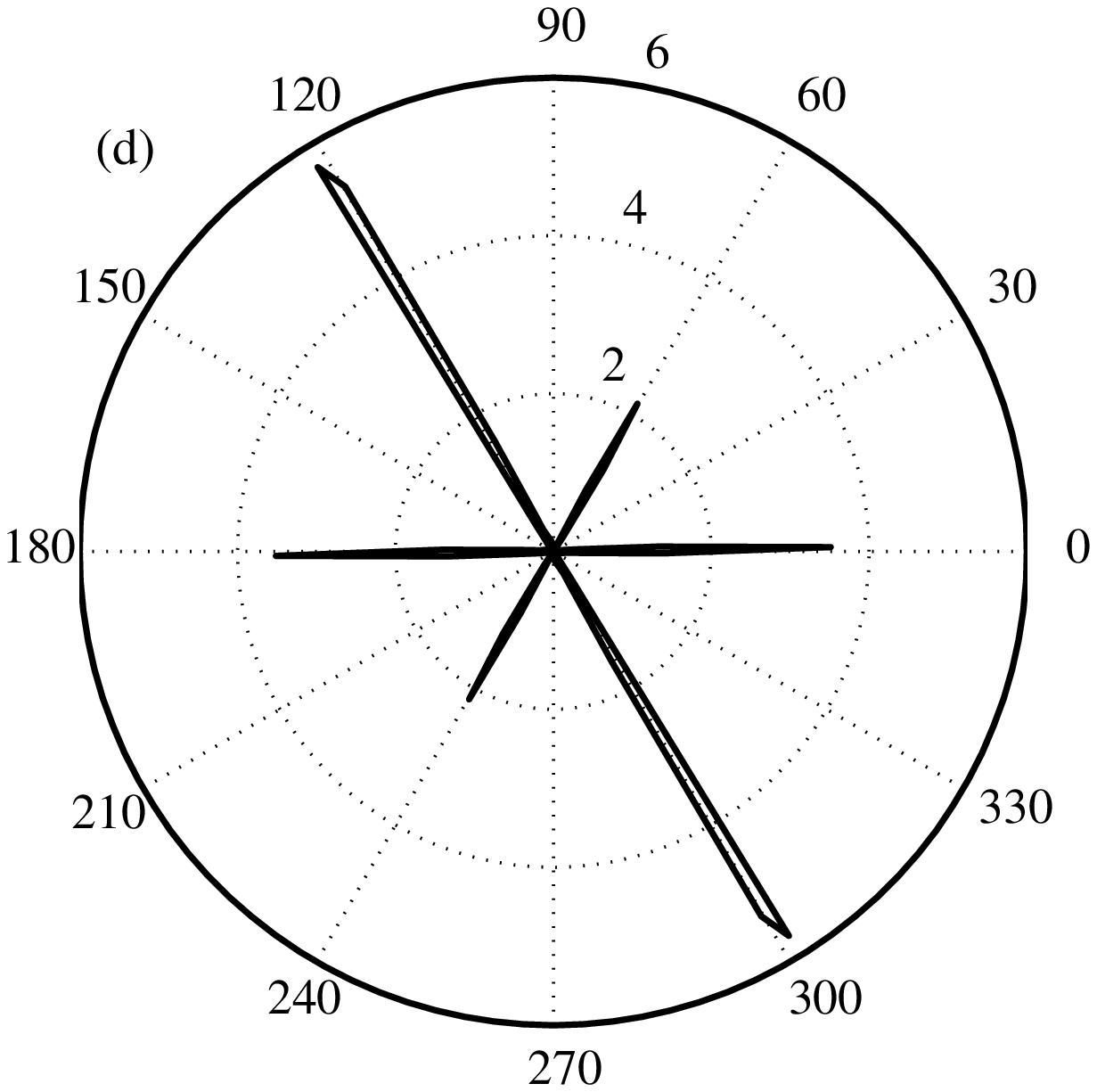,width=0.48\textwidth,angle=-00}}
\end{center}
\caption{
Evidence of {\it crystallization} in a sheared dense granular fluid at
$\nu=0.8$ and $e=0.9$.
(a) Particle distribution and
(b) streamwise velocity at  $t=390$;
(c) evolution of streamwise velocity at early times;
(d) collision angle distribution.
}
\label{fig:fig_crystal1}
\end{figure}

\newpage
\begin{figure}[htbp]
\begin{center}
\parbox{0.48\textwidth}{\epsfig{file=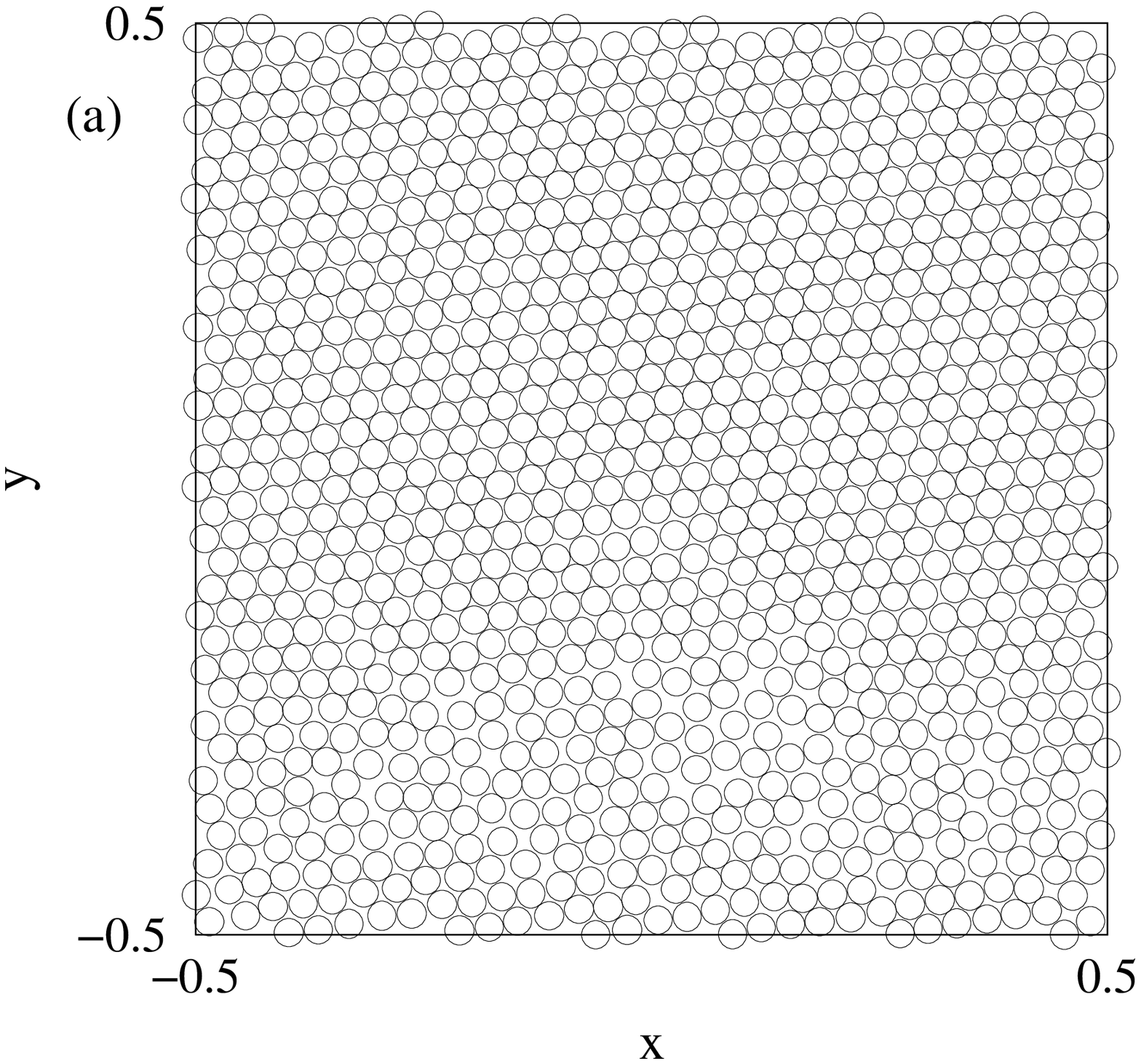,width=0.48\textwidth,angle=-00}}
\parbox{0.48\textwidth}{\epsfig{file=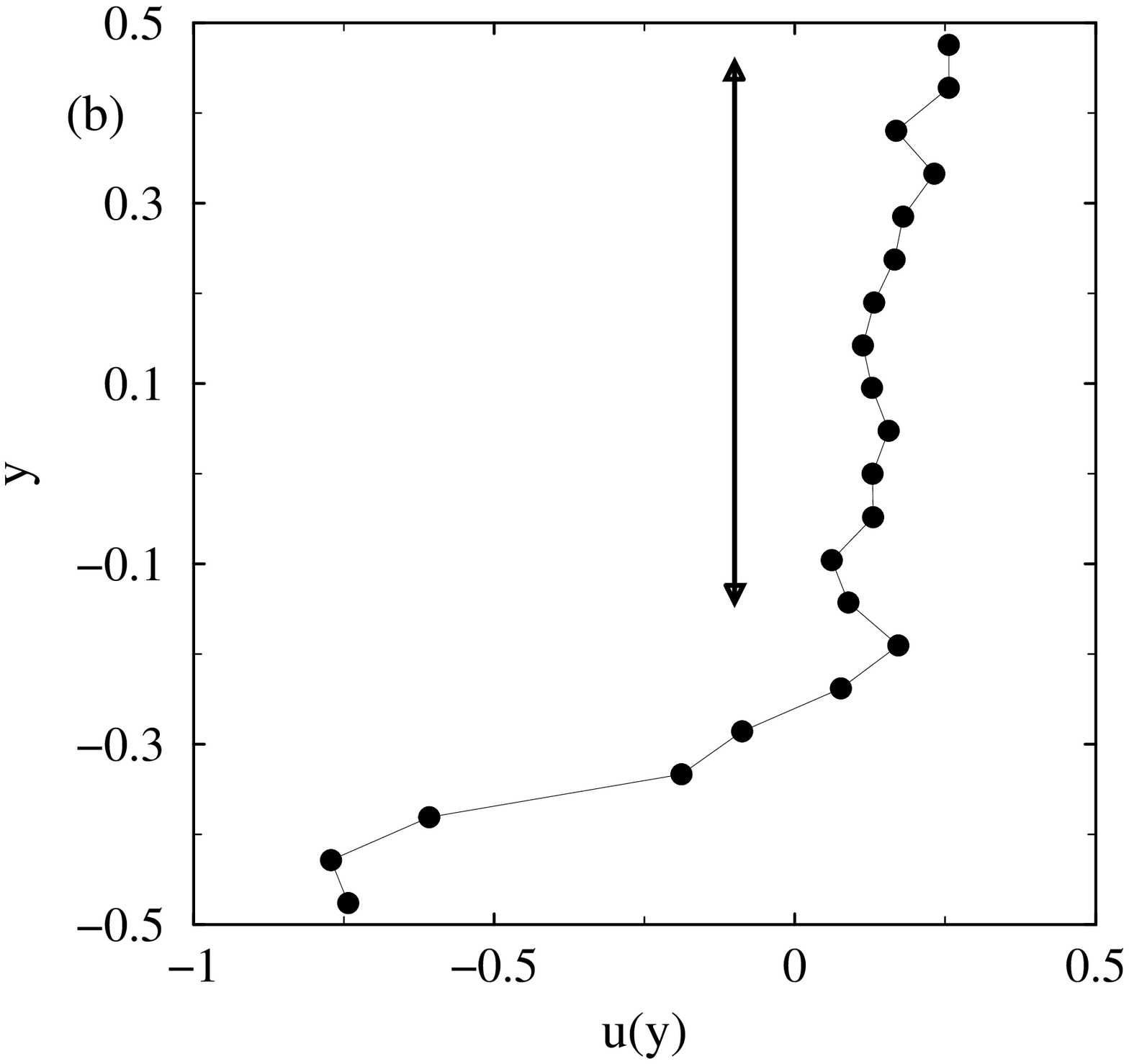,width=0.48\textwidth,angle=-00}}\\
\parbox{0.48\textwidth}{\epsfig{file=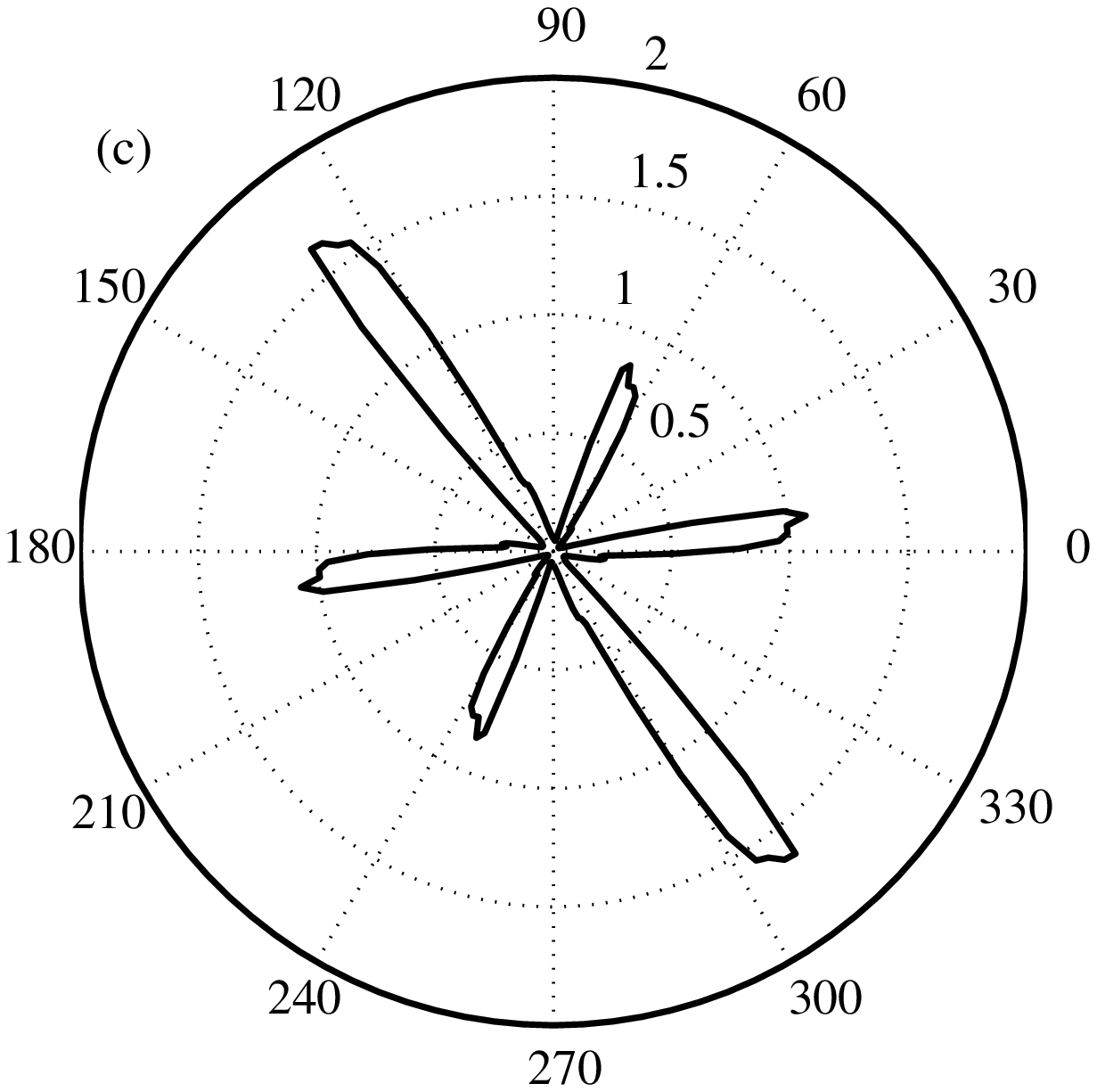,width=0.48\textwidth,angle=-00}}
\end{center}
\caption{
Effect of the coefficient of restitution on
{\it crystallization}:
$\nu=0.8$ and $e=0.99$.
(a) Particle distribution and
(b) streamwise velocity at $t=150$; (c) collision angle distribution.
}
\label{fig:fig_crystal2}
\end{figure}

\newpage
\begin{figure}[htbp]
\begin{center}
\parbox{0.75\textwidth}{\epsfig{file=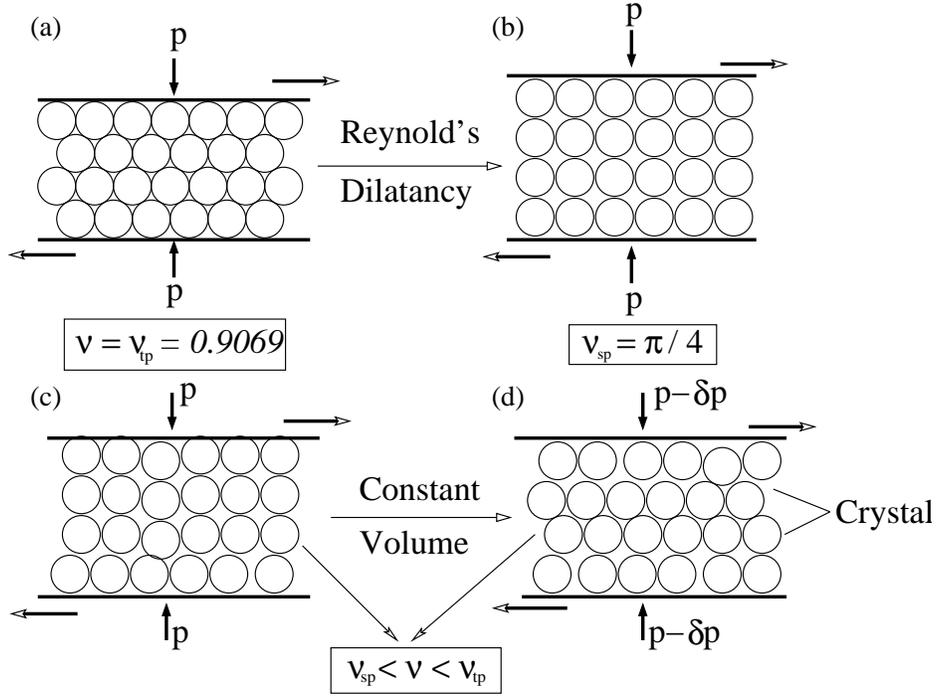,width=0.75\textwidth,angle=-00}}
\end{center}
\caption{
 A schematic diagram to explain Reynold's dilatancy in the Couette shear flow.
}
\label{fig:fig_reynolds}
\end{figure}

\newpage
\begin{figure}[htbp]
\begin{center}
\parbox{0.62\textwidth}{\epsfig{file=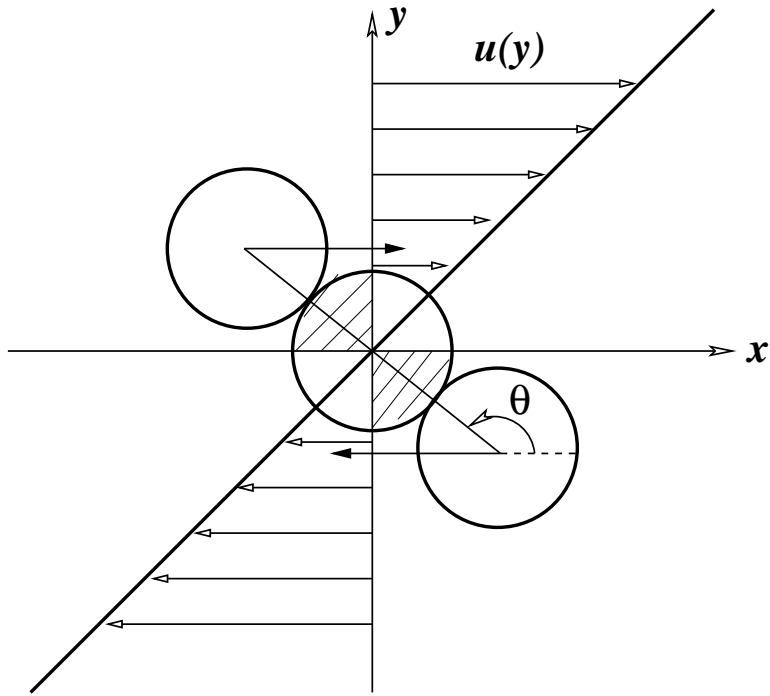,width=0.62\textwidth,angle=-00}}
\end{center}
\caption{
A schematic of the possible collision angles $\theta$ in uniform shear flow;
$\theta$ is measured anticlockwise from the positive $x$-axis.
Note that the collisions are more likely to occur in the second
($\pi/2<\theta<\pi$) and fourth ($3\pi/2<\theta<2\pi$) quadrants of the colliding disks.
}
\label{fig:fig_appendix1}
\end{figure}

\end{document}